\documentclass[11pt,a4paper]{article}
\usepackage{latexsym}
\usepackage{epsfig}
\usepackage{amsmath}
\usepackage{amssymb}
\usepackage{color}
\usepackage{bm}
\usepackage{amsthm}
\usepackage{amsthm,ctable}
\usepackage[dvipdfm,CJKbookmarks,bookmarksopen=true,colorlinks=true,
linkcolor=blue,citecolor=blue,pdfstartview=FitH,pdftitle=title,pdfauthor=lixx]{hyperref}
\RequirePackage[mathlines, displaymath]{lineno}      

\newcommand{\be}{\begin{equation}}
\newcommand{\ee}{\end{equation}}
\newcommand{\bea}{\begin{eqnarray}}
\newcommand{\eea}{\end{eqnarray}}
\newcommand{\beas}{\begin{eqnarray*}}
\newcommand{\eeas}{\end{eqnarray*}}
\newtheorem{lemma}{{\bf Lemma}}
\newtheorem{remark}{{\bf Remark}}
\newtheorem{theorem}{{\bf Theorem}}

\def\Xi{X^{(i)}}

\makeatletter      
\@addtoreset{equation}{section}
\makeatother       

\begin{document}


\title{Simultaneous Confidence Band for
 Partially Linear Panel Data Models with Fixed Effects\footnotetext{
Supported by the National Natural Science Foundation of China (No. 11471029),
the Beijing Natural Science Foundation (No. 1142002) and the  Science and
Technology Project  of Beijing Municipal Education Commission (No.
KM201410005010). } }

\author{Xiujuan YANG$^{1}$, Suigen YANG$^{2}$ and Gaorong LI$^{3}$
\thanks{
Gaorong Li is the corresponding author. E-mail: ligaorong@gmail.com}\\
\\
{\small $^{1}$College of Applied Sciences, Beijing University of Technology, Beijing 100124, China}\\
{\small $^{2}$College of Sciences, Tianjin University of Commerce, Tianjin 300134, China}\\
{\small $^{3}$Beijing Institute for Scientific and Engineering Computing, }\\
{\small Beijing University of Technology, Beijing 100124, China}\\
 }

\date{}
\maketitle

\begin{abstract}
In this paper, we construct the simultaneous confidence band (SCB)
for the nonparametric component in partially linear panel data
models with fixed effects. We remove the fixed effects, and further
obtain the estimators of parametric and nonparametric components,
which do not depend on the fixed effects. We establish the
asymptotic distribution of their maximum absolute deviation between
the estimated nonparametric component and the true nonparametric
component under some suitable conditions, and hence the result can
be used to construct the simultaneous confidence band of the
nonparametric component. Based on the asymptotic distribution,   it
becomes difficult for the construction of the simultaneous
confidence band. The reason is that the asymptotic distribution
involves the estimators of the asymptotic bias and conditional
variance, and the choice of the bandwidth for estimating the second
derivative of nonparametric function. Clearly, these will cause
computational burden and accumulative errors. To overcome these
problems, we propose a Bootstrap method to construct simultaneous
confidence band. Simulation studies indicate that the proposed
Bootstrap method exhibits better performance under the limited
samples.

\end{abstract}

\noindent {\footnotesize {\it 2000 MR Subject Classification:}
62G08; 62H12;  62G20}


\section{Introduction}

In the literature, there were a large amount of studies about
parametric linear and nonlinear  panel data models, and Arellano
(2003), Baltigi (2005), and Hsiao (2003) had provided excellent
overview of parametric panel data model analysis. To relax the
strong restrictions assumed in the parametric panel data models,
nonparametric and  semiparametric panel data models have received a
lot of attention in recent years. Compared to traditional parametric
panel data model, nonparametric or semiparametric panel data models
are better and more flexible to fit the actual data. Thus, this kind
of models have become the hot research topic for the econometricians
and statisticians. For example, Henderson, Carroll and Li (2008),
and Li, Peng and Tong (2013) considered the fixed effects
nonparametric panel data model. Henderson and Ullah (2005), Lin and
Ying (2001), and Wu and Zhang (2002) considered the random effects
nonparametric panel data models. Li and Stengos (1996) considered a
partially linear panel data model with some regressors being
endogenous via IV approach. Su and Ullah (2006) investigated the
fixed effects partially linear panel data model with exogenous
regressors. Zhang et al. (2011) considered the empirical likelihood
inference for the fixed effects partially linear panel data model.
Sun, Carroll and Li (2009) considered the problem of estimating a
varying coefficient panel data model with fixed effects using a
local linear regression approach. Chen, Gao and Li (2013a, 2013b)
and Lai, Li and Lian (2013) studied the semiparametric estimation
for a single-index panel data model,    and among others.

Recently, the fixed effects models are frequently used in
econometrics and biometrics. In this paper, we consider the
following partially linear panel data models with fixed effects:

\begin{equation}\label{1-2}
Y_{it}=
\bm{X}_{it}^{\tau}\bm{\beta}+g(Z_{it})+\alpha_{i}+V_{it}, \quad
i=1,\cdots, n,\quad t=1,\cdots, T,
\end{equation}
where $\{\bm{X}_{it}\}$ are $p\times1$ vector of observable
regressors, $\{Z_{it}\}$ are explanatory variables in [0,1],
$\bm\beta$ is a $p\times1$ vector of unknown coefficients,
$g(\cdot)$ is an unknown smooth function in [0,1],  $\{V_{it}\}$ are
random errors with zero mean, and $\{\alpha_{i}\}$ are fixed
effects. In addition, $T$ is the time series length, $n$ is the
cross section size.

For model (\ref{1-2}), we assume that $\{\alpha_{i}\}$ are
unobserved time-invariant individual effects. Model (\ref{1-2}) is
called as a partially linear fixed effects model if $\{\alpha_{i}\}$
are correlated with $\{\bm{X}_{it},Z_{it}\}$ with an unknown
correlation structure. For identification purpose, we impose
$\sum\limits_{i=1}^{n}\alpha_{i}=0$. An application of fixed effects
models  is the study of individual wage rate, $\alpha_{i}$
represents different unobserved abilities of individual $i$, such as
the unmeasured skills or unobservable characteristics of individual
$i$, which maybe correlate with some observed covariates: age,
educational level, job grade, gender, work experience and {\it et
al.}. As a special case, when $\{\alpha_{i}\}$ are uncorrelated with
$\{\bm{X}_{it},Z_{it}\}$, model (\ref{1-2}) becomes a partially
linear random effects model.

  Baltagi and Li (2002) applied the first-order
difference to eliminate the fixed effects and used the series method
to estimate the parametric and nonparametric components, and they
further established the asymptotic properties. Su and Ullah (2006)
considered the estimation of partially linear panel data models with
fixed effects. Zhang et al. (2011) applied the empirical likelihood
method  to model (\ref{1-2}).

For the partially linear panel data models, the existing literatures
 considered the pointwise asymptotic normality of the estimator for
 the nonparametric component, and the result can be used to
construct the pointwise confidence bands. In practice,  we need to
construct the simultaneous confidence band  of the nonparametric
function in the model. The simultaneous confidence band is a
powerful tool  to check the graphical representation of the
nonparametric function during the practical applications. Therefore,
there are  extensive literatures on the construction of the
simultaneous confidence band. For example, Fan and Zhang (2000), and
Zhang and Peng (2010) considered the simultaneous confidence bands
for the coefficient functions in varying-coefficient models; Li,
Peng and Tong (2013) considered the simultaneous confidence band for
nonparametric fixed effects panel data model; Li et al. (2014) and Yang et al. (2014) studied the simultaneous confidence band and hypothesis
testing for the link function in single-index models, and more
literatures see Yothers and Sampson (2011), Brabanter et al. (2012),
Cao et al. (2012),  Liu et al. (2013), and Li and Yang (2015).

In this paper, combining the idea of least-squares dummy-variable
approach in parametric panel data  models with the local linear
regression technique in  nonparametric models, we use the profile
least-squares dummy-variable method proposed in Su and Ullah (2006)
to remove the fixed effects, and further obtain the estimators of
parametric  and nonparametric components, which do not depend on the
fixed effects. Under some suitable conditions, we establish the
asymptotic distribution of their maximum absolute deviation between
the estimated nonparametric component and the true nonparametric
component, and hence the result can be used to construct the
simultaneous confidence band of the nonparametric component. In
order to construct the simultaneous confidence band  based on the
asymptotic distribution, we first need to estimate the asymptotic
bias and conditional variance, and choose the bandwidth for
estimating the second derivative of nonparametric function. These
will cause computational burden and accumulative errors, and it
becomes difficult for the construction of the simultaneous
confidence band. To overcome these problems, we further propose a
Bootstrap method to construct the simultaneous confidence band of
the nonparametric component in model (\ref{1-2}).

The rest of the paper is organized as follows. In Section 2, we use
the profile least-squares dummy-variable approach to obtain the
estimators of the parametric and  nonparametric components, and
present the asymptotic properties. In Section 3, we propose the
Bootstrap method to construct the simultaneous confidence band. In
Section 4, simulation studies are used to illustrate the proposed
method under the limited samples. The technical proofs of the main
theorems are presented in the Appendix.

\section{Estimation procedure and asymptotic properties}

\subsection{Estimation procedure}

 Let $\{(Y_{it};\bm{X}_{it}^{\tau},Z_{it}),i=1,\cdots, n, t=1,\cdots,
T\}$  be an independent identically distributed (i.i.d.) random
sample which comes from model (\ref{1-2}). In this paper, we
consider the asymptotic theories by letting $n$ approach infinity
and holding $T$ fixed. In this section, we consider the estimation
procedure to first remove the fixed effects, and further obtain the
efficient estimators of  parametric and nonparametric components.

For ease of notation, let
\begin{eqnarray*}
&&\bm{Y}=(Y_{11},\cdots,Y_{1T},
Y_{21},\cdots,Y_{2T},\cdots,Y_{n1},\cdots,Y_{nT})^{\tau},\\
&&\bm{g}=\Big(g(Z_{11}),\cdots, g(Z_{1T}),
g(Z_{21}),\cdots,g(Z_{2T}),
\cdots ,g(Z_{n1}), \cdots,g(Z_{nT})\Big)^{\tau},\\
&&\bm{V}=(V_{11},\cdots,V_{1T}, V_{21},\cdots,V_{2T},\cdots,
V_{n1},\cdots,V_{nT})^{\tau},\\
&&\bm{\alpha}_{0}=(\alpha_{1},\cdots,\alpha_{n})^{\tau}
\end{eqnarray*}
and $\mathbf{X}=(\bm{X}_{11},\cdots,\bm{X}_{1T},
\bm{X}_{21},\cdots,\bm{X}_{2T},\cdots,\bm{X}_{n1},\cdots,\bm{X}_{nT})^{\tau}$
is an $nT\times p$ matrix, where
$\bm{X}_{it}=(X_{it1},\cdots,X_{itp})^{\tau}$. Then model
(\ref{1-2}) can be written as the following matrix form,
\begin{eqnarray}\label{3-2}
\bm{Y}=\mathbf{X}\bm{\beta}+\bm{g}+(\mathbf{I}_{n}\otimes
\bm{e}_{T})\bm{\alpha}_{0}+\bm{V},
\end{eqnarray}
where $\mathbf{I}_{n}$ is an $n\times n$ identity matrix,
$\bm{e}_{T}$ is a $T$-dimensional column vector with all elements
being 1, and $\otimes$ denotes the Kronecker product. Furthermore,
by the identification assumption
$\sum\limits_{i=1}^{n}\alpha_{i}=0$, we have
$\alpha_{1}=-\sum\limits_{i=2}^{n}\alpha_{i}$. Define the
$(nT)\times(n-1)$ matrix
$\mathbf{D}=[-\bm{e}_{n-1},\mathbf{I}_{n-1}]^{\tau}\otimes
\bm{e}_{T}$, and
$\bm{\alpha}=(\alpha_{2},\cdots,\alpha_{n})^{\tau}$, model
(\ref{3-2}) can be rewritten as
\begin{eqnarray}\label{3-3}
\bm{Y}=\mathbf{X}\bm{\beta}+\bm{g}+\mathbf{D}\bm{\alpha}+\bm{V}.
\end{eqnarray}

Given $\bm{\alpha}$ and $\bm{\beta}$, model (\ref{3-3}) is a version
of the usual nonparametric fixed effects panel data model
\begin{eqnarray}\label{3-4}
\bm{Y}-\mathbf{X}\bm{\beta}-\mathbf{D}\bm{\alpha}=\bm{g}+\bm{V}.
\end{eqnarray}
We first apply the local polynomial method (see the details in Fan
and Gijbels, 1996) to estimate the nonparametric function
$g(\cdot)$. For $Z_{it}$ in a small neighborhood of $z\in[0,1]$,
approximate $g(Z_{it})$ by
\begin{eqnarray}\label{3-5}
g(Z_{it})\approx g(z)+g'(z)(Z_{it}-z).
\end{eqnarray}
\par

Let $K(\cdot)$ is a kernel function in $\mathbb{R}$,
$K_{h}(z)=K(z/h)/h$, where $h$ is a bandwidth, and let
$$\mathbf{Z}_{z}=\left (\begin{array}{ccccccc}1&\cdots&1&\cdots&1&\cdots&1 \\
    Z_{11}-z&\cdots&Z_{1T}-z&\cdots&Z_{n1}-z&\cdots&Z_{nT}-z
            \end{array} \right )^{\tau},$$
$\mathbf{W}_{z}=\mathrm{diag}(K_{h}(Z_{11}-z),\cdots,K_{h}(Z_{1T}-z),
K_{h}(Z_{21}-z),\cdots,K_{h}(Z_{2T}-z),\cdots,K_{h}(Z_{n1}-z),\cdots,
K_{h}(Z_{nT}-z))$ is an $(nT)\times(nT)$ diagonal matrix. Let
$\bm{G}(z)=(g(z),(g'(z)))^{\tau}$,
$\bm{\eta}=(\bm{\alpha}^{\tau},\bm{\beta}^{\tau})^{\tau}$.
\par

In what follows, we outline the  estimation procedure for
$\bm{\beta}$ and $g(\cdot)$.
\par
 Given  $\bm\eta=(\bm{\alpha}^{\tau}, \bm{\beta}^{\tau})^{\tau}$, we
 define the following weighted least-squares objective function
\begin{eqnarray}\label{3-6}
(\bm{Y}-\mathbf{X}\bm{\beta}-\mathbf{Z}_{z}\bm{G}(z)-\mathbf{D}\bm{\alpha})^{\tau}\mathbf{W}_{z}(\bm{Y}-\mathbf{X}\bm{\beta}-\mathbf{Z}_{z}\bm{G}(z)-\mathbf{D}\bm{\alpha}).
\end{eqnarray}
Minimizing the above objective function (\ref{3-6}) with respect to
$\bm{G}(z)$, we can obtain the solution of $\bm{G}(z)$ as follows
\begin{eqnarray}\label{3-7}
\widetilde{\bm{{G}}}(z,\bm{\eta})=(\mathbf{Z}_{z}^{\tau}\mathbf{W}_{z}\mathbf{Z}_{z})^{-1}\mathbf{Z}_{z}^{\tau}\mathbf{W}_{z}(\bm{Y}-\mathbf{X}\bm{\beta}-\mathbf{D}\bm{\alpha}).
\end{eqnarray}
Define the smoothing operator by
$$\mathbf{M}(z)=(\mathbf{Z}_{z}^{\tau}\mathbf{W}_{z}\mathbf{Z}_{z})^{-1}\mathbf{Z}_{z}^{\tau}\mathbf{W}_{z}.$$
Then, we can define the estimator of $g(z)$ by
\begin{eqnarray}\label{3-8}
\widetilde{g}(z,\bm{\eta})=\bm{m}^{\tau}(z)(\bm{Y}-\mathbf{X}\bm{\beta}-\mathbf{D}\bm{\alpha}),
\end{eqnarray}
where $\bm{m}^{\tau}(z)=\bm{e}^{\tau}\mathbf{M}(z)$,
$\bm{e}=(1,0)^{\tau}$ is a $2\times1$ vector.
\par
Since the fixed effects is an $n$-dimensional unobserved variable,
it is difficult to obtain the consistent estimator for the fixed
effects. Therefore, we first need to remove the fixed effects from
the model, and further obtain the estimators of parametric and
nonparametric components. By (\ref{3-8}), we define the following
objective function
\begin{eqnarray}\nonumber
&&(\bm{Y}-\mathbf{X}\bm{\beta}-\widetilde{\bm
g}_{\bm\eta}(z)-\mathbf{D}\bm{\alpha})^{\tau}(\bm{Y}-\mathbf{X}\bm{\beta}-\widetilde{\bm
g}_{\bm\eta}(z)-\mathbf{D}\bm{\alpha})\\\nonumber
&=&[\bm{Y}-\mathbf{X}\bm{\beta}-\mathbf{M}(\bm{Y}-\mathbf{X}\bm{\beta}-\mathbf{D}\bm{\alpha})-\mathbf{D}\bm{\alpha}]^{\tau}[\bm{Y}-\mathbf{X}\bm{\beta}-\mathbf{M}(\bm{Y}-\mathbf{X}\bm{\beta}-\mathbf{D}\bm{\alpha})-\mathbf{D}\bm{\alpha}]\\
&=&(\widetilde{\bm{Y}}-\widetilde{\mathbf{X}}\bm{\beta}-\widetilde{\mathbf{D}}\bm{\alpha})^{\tau}(\widetilde{\bm{Y}}-\widetilde{\mathbf{X}}\bm{\beta}-\widetilde{\mathbf{D}}\bm{\alpha})\label{3-9},
\end{eqnarray}
where $\widetilde{\bm
g}_{\bm\eta}(z)=(\widetilde{g}(Z_{11},\bm{\eta}),\cdots,\widetilde{g}(Z_{1T},\bm{\eta}),\cdots,\widetilde{g}(Z_{n1},\bm{\eta}),\cdots,\widetilde{g}(Z_{nT},\bm{\eta})$,
$\widetilde{\bm{Y}}=(\mathbf{I}_{nT}-\mathbf{M})\bm{Y}$,
$\widetilde{\mathbf{X}}=(\mathbf{I}_{nT}-\mathbf{M})\mathbf{X}$,
$\widetilde{\mathbf{D}}=(\mathbf{I}_{nT}-\mathbf{M})\mathbf{D}$,
$\widetilde{\mathbf{Q}}=\mathbf{I}_{nT}-\widetilde{\mathbf{D}}(\widetilde{\mathbf{D}}^{\tau}\widetilde{\mathbf{D}})^{-1}\widetilde{\mathbf{D}}^{\tau}$,
and $\mathbf{M}$ is an $(nT)\times(nT)$ smoothing matrix, that is
$$\mathbf{M}=\left(
                \begin{array}{c}
                  (1, 0)(\mathbf{Z}_{Z_{11}}^{\tau}\mathbf{W}_{Z_{11}}\mathbf{Z}_{Z_{11}})^{-1}\mathbf{Z}_{Z_{11}}^{\tau}\mathbf{W}_{Z_{11}}\\
                  \vdots \\
                  (1, 0)(\mathbf{Z}_{Z_{1T}}^{\tau}\mathbf{W}_{Z_{1T}}\mathbf{Z}_{Z_{1T}})^{-1}\mathbf{Z}_{Z_{1T}}^{\tau}\mathbf{W}_{Z_{1T}}\\
                  \vdots \\
                  (1, 0)(\mathbf{Z}_{Z_{nT}}^{\tau}\mathbf{W}_{Z_{nT}}\mathbf{Z}_{Z_{nT}})^{-1}\mathbf{Z}_{Z_{nT}}^{\tau}\mathbf{W}_{Z_{nT}}\\
                    \end{array}
            \right).
               $$
In addition, let
$\mathbf{P}=(\mathbf{I}_{nT}-\mathbf{M})^{\tau}(\mathbf{I}_{nT}-\mathbf{M})$
be an $(nT)\times(nT)$ matrix.

Taking derivative of (\ref{3-9}) with respect to $\bm\alpha$ and
setting it equal to zero, we have
\begin{eqnarray}\label{bu}
\bm{\widetilde{\alpha}}(\bm\beta)=(\widetilde{\mathbf{D}}^{\tau}\widetilde{\mathbf{D}})^{-1}\widetilde{\mathbf{D}}^{\tau}(\widetilde{\bm{Y}}-\widetilde{\mathbf{X}}\bm{\beta}).
\end{eqnarray}
Obviously,  the estimator of the fixed effects depends on
$\bm\beta$. Based on the idea of least-squares dummy-variable
approach in panel data parametric models and the nonparametric local
linear regression technique, we then apply the profile least-squares
dummy variable method to estimate  parameter vector $\bm\beta$.

Plugging (\ref{bu}) into (\ref{3-9}), we then minimize the profile
least-squares objective function with respect to $\bm\beta$. Thus,
we obtain the  profile least-squares estimator of $\bm\beta$ as
\begin{eqnarray}\label{3-10}
\bm{\hat{\beta}}=(\widetilde{\mathbf{X}}^{\tau}\widetilde{\mathbf{Q}}\widetilde{\mathbf{X}})^{-1}\widetilde{\mathbf{X}}^{\tau}\widetilde{\mathbf{Q}}\widetilde{\bm{Y}}.
\end{eqnarray}

By (\ref{3-10}) and (\ref{bu}), we have
\begin{eqnarray}\label{3-11}
\bm{\hat{\alpha}}=(\hat{\alpha}_{2},\cdots,\hat{\alpha}_{n})=(\widetilde{\mathbf{D}}^{\tau}\widetilde{\mathbf{D}})^{-1}\widetilde{\mathbf{D}}^{\tau}(\widetilde{\bm{Y}}-\widetilde{\mathbf{X}}\hat{\bm{\beta}}).
\end{eqnarray}
By $\sum\limits_{i=1}^n\alpha_i=0$ and (\ref{3-11}), the estimator
of $\alpha_{1}$ is
$\hat{\alpha}_{1}=-\sum\limits_{i=2}^{n}\hat{\alpha}_{i}$.

By (\ref{3-7}), (\ref{3-10}) and (\ref{3-11}), and  some simple
calculations,  we can obtain the estimator of $\bm{{G}}(z)$ as
follows
\begin{eqnarray}\nonumber
\bm{\hat{G}}(z)&=&\bm{\hat{G}}(z,
\hat{\bm{\eta}})=\mathbf{M}(z)(\bm{Y}-\mathbf{X}\hat{\bm{\beta}}-\mathbf{D}\hat{\bm{\alpha}})\\\nonumber
&=&\mathbf{M}(z)[\bm{Y}-\mathbf{X}\hat{\bm{\beta}}-\mathbf{D}(\widetilde{\mathbf{D}}^{\tau}\widetilde{\mathbf{D}})^{-1}\widetilde{\mathbf{D}}^{\tau}(\widetilde{\bm{Y}}-\widetilde{\mathbf{X}}\hat{\bm{\beta}})]\\
&=&\mathbf{M}(z)(\mathbf{I}_{nT}-\mathbf{D(D^{\tau}PD)}^{-1}\mathbf{D^{\tau}P})(\bm{Y}-\mathbf{X}\hat{\bm{\beta}}).\label{3-12}
\end{eqnarray}
By (\ref{3-8}) and (\ref{3-12}), we get the estimator of $g(z)$ as
\begin{eqnarray}\label{3-13}
\hat{g}(z)=\bm{m}^{\tau}(z)(\mathbf{I}_{nT}-\mathbf{D(D^{\tau}PD)}^{-1}\mathbf{D^{\tau}P})(\bm{Y}-\mathbf{X}\hat{\bm{\beta}}).
\end{eqnarray}

\par
{\remark\label{rem1} From (\ref{3-10}) and (\ref{3-13}), it is easy
to see that  the estimators of $\bm\beta$ and $g(\cdot)$ do not
depend on the fixed effects.}

\subsection{Asymptotic properties}

Let $\mu_l=\int z^{l}K(z)dz$ and $\nu_l=\int z^{l}K^{2}(z)dz$ for
$l=0,1,2$. Define the observed covariate set by
$\mathcal{D}=\{\bm{X}_{it}, Z_{it}, 1\leq i\leq n, 1\leq t\leq T\}$.
In order to obtain the main results, we first present  the following
technical conditions.

(C1)\ \
$(\alpha_{i},\bm{V}_{i},\mathbf{X}_{i},\bm{Z}_{i}),i=1,\cdots,n$,
are i.i.d., where $\bm{V}_{i}=({V}_{i1},
{V}_{i2},\cdots,{V}_{iT})^{\tau}$, and $\mathbf{X}_{i}$ and
$\bm{Z}_{i}$ can be defined similarly.
$E\|\bm{X}_{it}\|^{2+\delta}<\infty$ and $
E\|V_{it}\|^{2+\delta}<\infty$ for some $\delta>0$. Let
$\sigma^{2}(\bm{x},z)=\mathrm{Var}(Y_{it}|\bm{X}_{it}=\bm{x},Z_{it}=z)$,
$\sigma^{2}(z)=\mathrm{Var}(Y_{it}|Z_{it}=z)$, and
$0<\sigma^{2}(\bm{x},z),\sigma^{2}(z)<\infty$.

(C2)\ \
$E(Y_{it}|\mathbf{X}_{i},\bm{Z}_{i},\alpha_{i})=E(Y_{it}|\bm{X}_{it},Z_{it},\alpha_{i})=\bm{X}_{it}^{\tau}\bm{\beta}+g(Z_{it})+\alpha_{i},
i=1,\cdots,n, t=1,\cdots,T$.

(C3)\ \ Let $f(z)=\sum\limits_{t=1}^{T}f_{t}(z)$, where $f_{t}(z)$
is the continuous density function of $Z_{it}$, and $f_{t}(z)$ is
bounded away from zero and infinity on $[0,1]$ for each
$t=1,\cdots,T$. Let
$\widetilde{V}_{it}=V_{it}-\frac{1}{T}\sum_{s=1}^{T}V_{is}$,
$\sigma_{t}^{2}(z)=E[\widetilde{V}^{2}_{it}|Z_{it}=z]$ and
$\bar{\sigma}^{2}(z)=\sum_{t=1}^{T}\sigma_{t}^{2}(z)f(z)$.

(C4)\ \ Let $\bm{p}(z)=E(\bm{X}_{it}|Z_{it}=z)$. The functions
$g(\cdot)$ and $\bm{p}(\cdot)$ have the bounded and continuous
second derivatives on $[0,1]$.

(C5)\ \ The kernel function $K(\cdot)$ is a symmetric density
function, and is absolutely continuous  on its support set $[-A,A]$.

~~~~(C5a)\ \ \ \ $K(A)\neq 0$ or

~~~~(C5b)\ \ \ \ $K(A)=0$, $K(t)$ is absolutely continuous and
$K^{2}(t)$, $[K'(t)]^{2}$ are integrable on the $(-\infty,+\infty)$.

(C6)\ \ The bandwidth $h$ satisfies that $nh^{3}/\log
n\rightarrow\infty$, $nh^{5}\log n\rightarrow 0$, as
$n\rightarrow\infty$.

\par
{\theorem\label{theo1} Assume that conditions  (C1)--(C6) hold. Let
$b(z)=h^{2} \mu_{2}g''(z)/2$,
 $\Sigma_{g}=\nu_{0}\bar{\sigma}^{2}(z)f^{-2}(z)$, $\Sigma_{g'}=\nu_{2}\bar{\sigma}^{2}(z)/(f^{2}(z)\mu^{2}_{2})$,
 Then uniformly for $z\in[0,1]$, we have
\begin{eqnarray*}
\|\hat{\bm\beta}-\bm\beta\|=O_{p}(n^{-1/2})
\end{eqnarray*}
and
$$
\sqrt{nh}\{\hat{g}(z)-g(z)-b(z)\}\stackrel{L}\longrightarrow
N(0,\Sigma_g),
$$
$$
\sqrt{nh^{3}}\{\hat{g}'(z)-g'(z)\}\stackrel{L}\longrightarrow
N(0,\Sigma_{g'}),
$$
where ``$\stackrel{L}\longrightarrow$" denotes the convergence in
distribution. }

\par
{\theorem\label{theo2} Assume that conditions  (C1)--(C6) hold and
$h=O(n^{-\rho})$ for $1/5\leq\rho<1/3$. Then for all $z\in[0,1]$, we
have
\begin{eqnarray*} &&P\left\{(-2\log
h)^{1/2}\Big(\sup\limits_{z\in[0,1]}\left|(nh\Sigma_{g}^{-1})^{1/2}(\hat{g}(z)-g(z)-b(z))\right|-d_{n}\Big)<u\right\}\\
&&~~~~~~\longrightarrow \exp \left(-2\exp(-u)\right),\quad \hbox{as
$n\rightarrow\infty$},
\end{eqnarray*}
where if $K(A)\neq 0$,
$$d_{n}=(-2\log h)^{1/2}+\frac{1}{(-2\log h)^{1/2}}\left\{\log\frac{K^2(A)}{\nu_0\pi^{1/2}}+\frac{1}{2}\log\log h^{-1}\right\},$$
and if $K(A)=0$,
$$d_{n}=(-2\log h)^{1/2}+\frac{1}{(-2\log h)^{1/2}}\log\left\{\frac{1}{4\nu_0\pi}\int(K'(z))^2dz\right\}.$$
 }

 Theorem \ref{theo2} gives the asymptotic distribution
of the maximum absolute deviation between the estimated
nonparametric component $\hat{g}(\cdot)$ and the true nonparametric
component $g(\cdot)$ when the estimator of $\bm\beta$ is
$\sqrt{n}-$consistent. It provides us the theoretical foundation for
constructing the simultaneous confidence band  of the nonparametric
function in  model (\ref{1-2}).

{\remark\label{rem2} If the supremum in Theorem \ref{theo2} is taken
on an interval of $[c,d]$ instead of $[0,1]$, Theorem \ref{theo2}
still holds under certain conditions by transformation. The
 asymptotic distribution is represented as
\begin{eqnarray*}
&&P\left\{(-2\log
h/(d-c))^{1/2}\Big(\sup\limits_{z\in[c,d]}\left|(nh\Sigma_{g}^{-1})^{1/2}(\hat{g}(z)-g(z)-b(z))\right|-\widetilde{d}_{n}\Big)<u\right\}\\
&&~~~~~~~~~~~~~~~~~\longrightarrow \exp
\left(-2\exp(-u)\right),
\end{eqnarray*}
where $\widetilde{d}_{n}$ is the same as $d_{n}$ in the Theorem
\ref{theo2} except that $h$ is replaced by $h/(d-c)$.}

\par
{\theorem\label{theo3} Assume that conditions (C1)--(C6) hold and
$\Sigma_{g'}=\nu_{2}\bar{\sigma}^{2}(z)/(f^{2}(z)\mu^{2}_{2})$. Then
for all $z\in[0,1]$, we have
\begin{eqnarray*} &&P\left\{(-2\log
h)^{1/2}\Big(\sup\limits_{z\in[0,1]}\left|(nh^3\Sigma_{g'}^{-1})^{1/2}(\hat{g}'(z)-g'(z))\right|-d_{n_{1}}\Big)<u\right\}\\
&&~~~~~~~~~~\longrightarrow \exp \left(-2\exp(-u)\right), \quad
\hbox{as $n\rightarrow\infty$},
\end{eqnarray*}
where $d_{n_{1}}=(-2\log h)^{1/2}+\frac{1}{(-2\log
h)^{1/2}}\log\left\{\frac{1}{2\pi\sqrt{\nu_2}}(\int
z^{2}(K'(z))^2dz)^{1/2}\right\}.$ If $K(c_{0})=0$, $K(z)$ is
absolutely continuous and $K^{2}(z)$, $(K'(z))^{2}$ are integrable
on $(-\infty, +\infty)$}.

Theorem \ref{theo3}  presents the asymptotic distribution of the
maximum absolute deviation for $\hat{g}'(\cdot)$

\subsection{Simultaneous confidence band for the nonparametric
function}

Since the asymptotic bias and variance of $\hat{g}(\cdot)$ in
Theorem \ref{theo2} involve some unknown quantities, we cannot apply
Theorem \ref{theo2} to construct simultaneous confidence band of
$g(\cdot)$ directly. In order to construct the simultaneous
confidence band of $g(\cdot)$, we first need to get the consistent
estimators of the asymptotic bias and variance of $\hat{g}(\cdot)$.
 By Theorem
\ref{theo1}, the asymptotic bias of $\hat{g}(z)$ is
$$(h^{2}\mu_{2}/2)g''(z)(1+o_{p}(1)).$$
Thus, the consistent estimator of the asymptotic bias  is
$\mathrm{\widehat{bias}}(\hat{g}(z))=h^{2}\mu_{2}\hat{g}''(z)/2$,
where the estimator  $\hat{g}''(z)$ of $g''(z)$ is obtained by using
local cubic fit with  an appropriate pilot bandwidth
$h_{*}=O(n^{-1/7})$, which is optimal for estimating $g''(z)$ and
can be chosen by the residual squares criterion proposed in Fan and
Gijbels (1996).

Next we will estimate the asymptotic variance of $\hat{g}(z)$. For
simplicity, suppose that the random errors $V_{it}$ are  i.i.d. for
all $i$ and $t$. By the proofs of theorem, we have
$$\mathrm{Var}\{\hat{g}(z)|\mathcal{D}\}=(1,0)(\mathbf{Z}_{z}^{\tau}\mathbf{W}_{z}\mathbf{Z}_{z})^{-1}(\mathbf{Z}_{z}^{\tau}\mathbf{W}_{z}\mathbf{Q}_{1}\Phi_{1}
\mathbf{Q}_{1}\mathbf{W}_{z}\mathbf{Z}_{z})(\mathbf{Z}_{z}^{\tau}\mathbf{W}_{z}\mathbf{Z}_{z})^{-1}(1,0)^{\tau},$$
where
$\mathbf{Q}_{1}=(\mathbf{I}_{nT}-\mathbf{D(D^{\tau}PD)}^{-1}\mathbf{D^{\tau}P})$
and
$\Phi_{1}=\mathrm{diag}(\sigma^{2}(Z_{11}),\cdots,\sigma^{2}(Z_{1T}),
\sigma^{2}(Z_{21}),\cdots,\\ \sigma^{2}(Z_{2T}),\cdots,
\sigma^{2}(Z_{n1}),\cdots,\sigma^{2}(Z_{nT}))$. Using the similar
approximate local  homoscedasticity in Li, Peng and Tong (2013), the
asymptotic variance of $\hat{g}(z)$ is defined by
$$\mathrm{Var}\{\hat{g}(z)|\mathcal{D}\}=(1,0)(\mathbf{Z}_{z}^{\tau}\mathbf{W}_{z}\mathbf{Z}_{z})^{-1}(\mathbf{Z}_{z}^{\tau}
\mathbf{W}_{z}\mathbf{Q}_{1}\mathbf{W}_{z}\mathbf{Z}_{z})(\mathbf{Z}_{z}^{\tau}\mathbf{W}_{z}\mathbf{Z}_{z})^{-1}(1,0)^{\tau}\sigma^{2}(z).$$

Let $\hat{\bm{V}}=\bm{Y}-\bm{\hat{Y}}$ be the residual, where
$\bm{\hat{Y}}=\bm{\hat{g}}+\bf{X}\hat{\bm{\beta}}+\bf{D}\hat{\bm{\alpha}}$.
By (\ref{3-10}), (\ref{3-11}) and (\ref{3-13}), we have
\begin{eqnarray}\nonumber
\hat{\bm{V}}&=&
\bm{Y}-\bm{\hat{g}}-\bf{X}\hat{\bm{\beta}}-\bf{D}\hat{\bm{\alpha}}\\\nonumber
              &=& \bm{Y}-\bf{X}\hat{\bm{\beta}}-\bf{D}\hat{\bm{\alpha}}-\mathbf{M}(\bm{Y}-\bf{X}\hat{\bm{\beta}}-\bf{D}\hat{\bm{\alpha}})\\\nonumber
              &=&(\mathbf{I}_{nT}-\mathbf{M})(\bm{Y}-\bf{X}\hat{\bm{\beta}}-\bf{D}\hat{\bm{\alpha}})\\\nonumber
              &=&(\mathbf{I}_{nT}-\mathbf{M})(\mathbf{I}_{nT}-\mathbf{D(D^{\tau}PD)}^{-1}\mathbf{D^{\tau}P})(\bm{Y}-\bf{X}\hat{\bm{\beta}})\\\nonumber
              &=&(\mathbf{I}_{nT}-\mathbf{M})\mathbf{Q}_{1}(\mathbf{I}_{nT}-\mathbf{X(X^{\tau}PQ}_{1}\mathbf{X})^{-1}\mathbf{X}^{\tau}\mathbf{PQ}_{1})\bm{Y}\\
              &=:&(\mathbf{I}_{nT}-\mathbf{M})\mathbf{Q}_{1}\mathbf{Q}_{2}\bm{Y},\label{cancha}
\end{eqnarray}
where
$\mathbf{Q}_{2}=\mathbf{I}_{nT}-\mathbf{X}(\mathbf{X}^{\tau}\mathbf{PQ}_{1}\mathbf{X})^{-1}\mathbf{X}^{\tau}\mathbf{PQ}_{1}$.
Obviously, the residual $\hat{\bm{V}}$ does not depend on the fixed
effects, and is a linear function of $\bm{Y}$. By  the normalized
weighted residual sum of squares,   $\sigma^{2}(z)$ can be estimated
by
$$\hat{\sigma}^{2}(z)=\frac{\hat{\bm{V}}^{\tau}\hat{\bm{V}}}{{\rm tr}(\mathbf{Q}_{2}^{\tau}\mathbf{Q}_{1}^{\tau}\mathbf{P}\mathbf{Q}_{1}\mathbf{Q}_{2})}
=\frac{\bm{Y}^{\tau}(\mathbf{Q}_{2}^{\tau}\mathbf{Q}_{1}^{\tau}\mathbf{P}\mathbf{Q}_{1}\mathbf{Q}_{2})\bm{Y}}{{\rm
tr}(\mathbf{Q}_{2}^{\tau}\mathbf{Q}_{1}^{\tau}\mathbf{P}\mathbf{Q}_{1}\mathbf{Q}_{2})}.
$$

{\theorem\label{theo4} Under the conditions in Theorem \ref{theo2},
and assume that $\hat{g}^{(3)}(\cdot)$ is continuous on $[0,1]$ and
the pilot bandwidth $h_{\ast}$ satisfies that
$h_{\ast}=O(n^{-1/7})$. Then  for all $z\in[0,1]$, we have
\begin{eqnarray*}
P\left\{(-2\log
h)^{1/2}\Big(\sup\limits_{z\in[0,1]}\left|\frac{\hat{g}(z)-g(z)-\widehat{\mathrm{bias}}(\hat{g}(z)|\mathcal{D})}{[\widehat{\mathrm{Var}}\{\hat{g}(z)|\mathcal{D}\}]^{1/2}}\right|-d_{n}\Big)<
u\right\} \longrightarrow \exp\left(-2\exp(-u)\right),
\end{eqnarray*}
where $d_{n}$ is defined in Theorem \ref{theo2}. }

\par
By Theorem \ref{theo4}, we construct the $(1-\alpha)\times100\%$
simultaneous confidence band of the nonparametric function  $g(z)$
as
\begin{eqnarray}\label{3-100}
\left(\hat{g}(z)-\widehat{\mathrm{bias}}(\hat{g}(z)|\mathcal{D})\pm\Delta_{1,\alpha}(z)\right),
\end{eqnarray}
where
$\Delta_{1,\alpha}(z)=\left(d_{n}+[\log2-\log\{-\log(1-\alpha)\}](-2\log
h)^{-1/2}\right)\left[\widehat{\mathrm{Var}}\{\hat{g}(z)|\mathcal{D}\}\right]^{1/2}.$

\section{The Bootstrap method}

Despite the fact that Theorem \ref{theo4} provides  the asymptotic
distribution   to construct the simultaneous confidence band
(\ref{3-100}) for the nonparametric component, we need to estimate
the asymptotic bias and the asymptotic conditional variance. First,
the estimator of the asymptotic bias involves the estimator the
second derivative $g''(\cdot)$ and the choice of the pilot bandwidth
$h_*$ for estimating the second derivative $g''(\cdot)$. The
estimator of the second derivative $g''(\cdot)$ has a slow
convergence rate, and is very sensitive with the pilot bandwidth
$h_*$. This will influence the estimator of the asymptotic bias.
Second, the asymptotic variance estimation is very complicated,
especially for panel data semiparametric  fixed effects model.
Finally, the asymptotic critical value $c_{\alpha}$ depends on the
double exponential distribution, the estimators of asymptotic bias
and the asymptotic conditional variance. These will not only cause
computational burden and accumulative errors, but also lead to the
difficulty to construct simultaneous confidence band. To overcome
these problems, we extend the Bootstrap method used in Li, Peng and
Tong (2013) to partially linear panel data fixed effects model
(\ref{1-2}).

Now we discuss how to use the Bootstrap procedure to construct
simultaneous confidence band for $g(\cdot)$. Let
$$T=\sup\limits_{z\in[0,1]}\frac{|\hat{g}(z)-g(z)|}{\{\mathrm{Var}(\hat{g}(z|\mathcal{D}))\}^{1/2}}.$$
Suppose that the upper $\alpha$ quantile of $T$ is $c_{\alpha}$. If
$c_{\alpha}$ and $\mathrm{Var}(\hat{g}(z|\mathcal{D}))$ are known,
the simultaneous confidence band of $g(\cdot)$ with
$(1-\alpha)\times100\%$ on the interval $[0,1]$ should be
$$\hat{g}(z)\pm\{\mathrm{Var}(\hat{g}(z|\mathcal{D}))\}^{1/2}c_{\alpha}.$$
However, $c_{\alpha}$ and $\mathrm{Var}(\hat{g}(z|\mathcal{D}))$ are
unknown. We will get their estimators using the bootstrap method.
Suppose that we have the estimators   $\hat{c}_{\alpha}$ and
$\mathrm{Var}^{\ast}(\hat{g}(z|\mathcal{D}))$ of $c_{\alpha}$ and
$\mathrm{Var}(\hat{g}(z|\mathcal{D}))$, respectively. Then we can
obtain the $(1-\alpha)\times100\%$ simultaneous confidence band of
$g(\cdot)$ as follows
\begin{eqnarray}\label{SCB-3}
\hat{g}(z)\pm\{\mathrm{Var}^{\ast}(\hat{g}(z|\mathcal{D}))\}^{1/2}\hat{c}_{\alpha}.
\end{eqnarray}
\par
The Bootstrap procedure is given as follows:

\par
(1)~~By (\ref{cancha}), obtain the residuals
$\bm{\hat{V}}=(\mathbf{I}_{nT}-\mathbf{M})\mathbf{Q}_{1}\mathbf{Q}_{2}\bm{Y}$,
where $\bm{\hat{V}}=(\hat{V}_{11},\cdots,\hat{V}_{1T},\\
\hat{V}_{21},\cdots, \hat{V}_{2T},
\cdots,\hat{V}_{n1},\cdots,\hat{V}_{nT})^{\tau}$.
\par
(2)~~For each $i=1,\cdots,n,~t=1,\cdots,T$, obtain the bootstrap
error $V^{\ast}_{it}=\hat{V}_{it}\varepsilon_{it}$, where
$\varepsilon_{it}$ are i.i.d. $\thicksim N(0,1)$ across $i$ and $t$.
Generate the bootstrap sample member $Y^{\ast}_{it}$ by
$Y^{\ast}_{it}=\hat{Y}_{it}+V^{\ast}_{it},~
i=1,\cdots,n,~t=1,\cdots,T$.
\par
(3)~~Given the bootstrap resample
$\{(Y^{\ast}_{it},\bm{X}_{it},Z_{it}),i=1,\cdots,n,~t=1,\cdots,T\}$,
obtain the estimators of $\bm{\beta}$ and $g(\cdot)$, and denote the
resulting estimate by $\bm{\hat{\beta}}^{\ast}$ and
$\hat{g}^{\ast}(\cdot)$, as the bootstrap estimators of $\bm{\beta}$
and $g(\cdot)$, respectively.
\par
(4)~~Repeat (2)--(3)~$N$ times to get a size $N$ bootstrap sample of
$g(\cdot)$, $\hat{g}^{\ast}_{k}(\cdot),k=1,\cdots,N$. The estimator
$\mathrm{Var}^{\ast}(\hat{g}(z))$ of $\mathrm{Var}(\hat{g}(\cdot))$
is taken as the sample variance of $\hat{g}^{\ast}_{k}(\cdot)$.
\par
(5)~~Compute the bootstrap sample of $T$ by
$$T^{\ast}_{k}=\sup\limits_{z\in[0,1]}\frac{|\hat{g}_{k}^{\ast}(z)-\hat{g}(z)|}{\{\mathrm{Var}^{\ast}(\hat{g}(z|\mathcal{D}))\}^{1/2}},
\qquad k=1,\cdots,N.$$ Use the upper $\alpha$ percentile
$\hat{c}_{\alpha}$ of $T^{\ast}_{k},k=1,\cdots,N,$
 to estimate the upper $\alpha$ quantile $c_{\alpha}$ of
$T$.
\par

We can construct the $(1-\alpha)\times100\%$ simultaneous confidence
band of $g(\cdot)$ by (\ref{SCB-3}) when we obtain the estimators of
$c_{\alpha}$ and $\mathrm{Var}(\hat{g}(z|\mathcal{D}))$.

\section{Simulation studies}

We conduct simulation studies to assess the performance of our
proposed method. Our simulated data are generated from the following
model:
\begin{eqnarray}\label{3-14}
Y_{it}=\bm{X}^{\tau}_{it}\bm{\beta}+0.8\cos(\pi
Z_{it})+\alpha_{i}+V_{it},~~ i=1,\cdots,n,
 ~~t=1,\cdots,T,
\end{eqnarray}
where $\bm{\beta}=(-1, 3, 5)^{\tau}$, $\bm{X}_{it}$ are three
dimensional i.i.d. random variables from uniform [-1,1], $Z_{it}$
are i.i.d. from uniform [-1,1], and the random errors $V_{it} $ are
i.i.d. from $N(0,1)$. In this simulation, we only consider
$\alpha_i$ are correlated with the covariate $Z_{i.}$, and generate
$\alpha_{i}=\varepsilon_{i}+cZ_{i\cdot}, i=2,\cdots,n$,
 where $\varepsilon_{i}
\thicksim N(0,1), Z_{i\cdot}=\frac{1}{T}\sum\limits_{t=1}^{T}Z_{it}$
and $\alpha_{1}=-\sum\limits^{n}_{i=2}\alpha_{i}, i=1,\cdots,n$. We
consider three cases for $c=0,0.5,1$. When $c\neq0$, $Z_{it}$ and
$\alpha_{i}$ are correlated, model (\ref{3-14}) is the partially
linear fixed effects model. When $c=0$, model (\ref{3-14}) leads to
the usual partially linear random effects model.
\par

In our simulation studies, we apply the Epanechnikov kernel
$K(z)=0.75(1-z^{2})_+$ for estimating the nonparametric function.
Finding an appropriate bandwidth can be of both theoretical and
practical interest. To implement the estimation procedure described
in Section 2, we need to choose the bandwidth $h$. One can select
$h$ by minimizing the generalized cross validation criterion. Here
we use the following
 cross validation method to automatically select the optimal bandwidth $h_{\rm CV}$.
\begin{eqnarray}\label{3-15}
{\rm
CV}(h)=\sum_{i=1}^{n}\sum_{t=1}^{T}(Y_{it}-\hat{Y}_{it}^{-it})^{2}
=\sum_{i=1}^{n}\sum_{t=1}^{T}\left(\frac{Y_{it}-\hat{Y}_{it}}{1-l_{kk}}\right)^2
=\sum_{i=1}^{n}\sum_{t=1}^{T}\left(\frac{\hat{V}_{it}}{1-l_{kk}}\right)^2,
\end{eqnarray}
where $Y_{it}^{-it}$ denote the fitted values that are computed from
data with measurements of  the $\{Y_{it}, {\bm X}_{it}\}$
observation deleted. $k=(i-1)T+t$,
$\hat{V}_{it}=Y_{it}-\hat{Y}_{it}$ and $l_{kk}$ is the ${(k, k)}$
element  of matrix
$[\mathbf{I}_{nT}-(\mathbf{I}_{nT}-\mathbf{M})\mathbf{Q}_{1}\mathbf{Q}_{2}]$.
The cross validation bandwidth $h_{\rm CV}$ is then defined to be
the minimizer of ${\rm CV}(h)$.
\par

\begin{figure}[htbp]
    \centering
    \includegraphics[width=4cm,height=6cm]{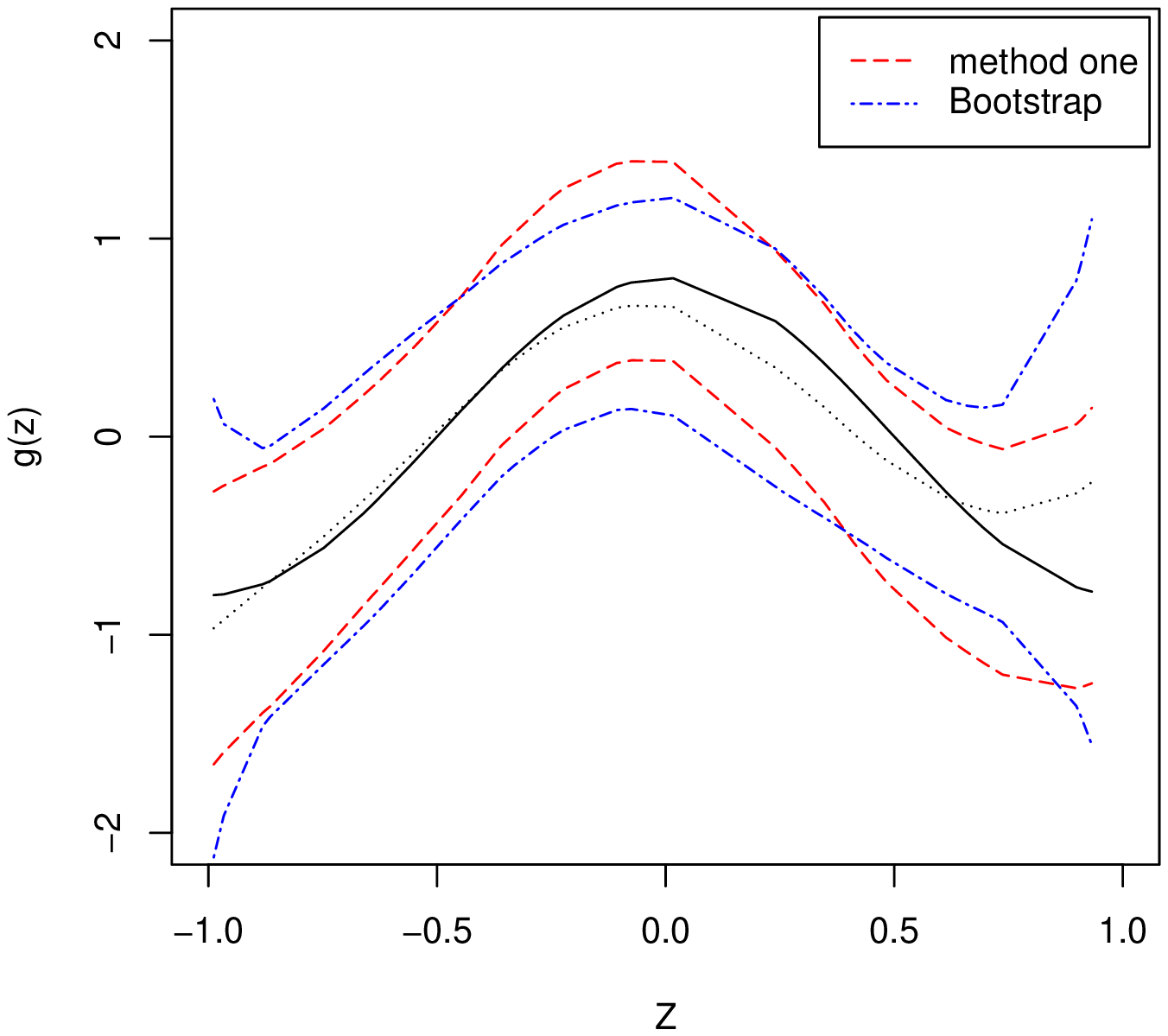}\includegraphics[width=4cm,height=6cm]{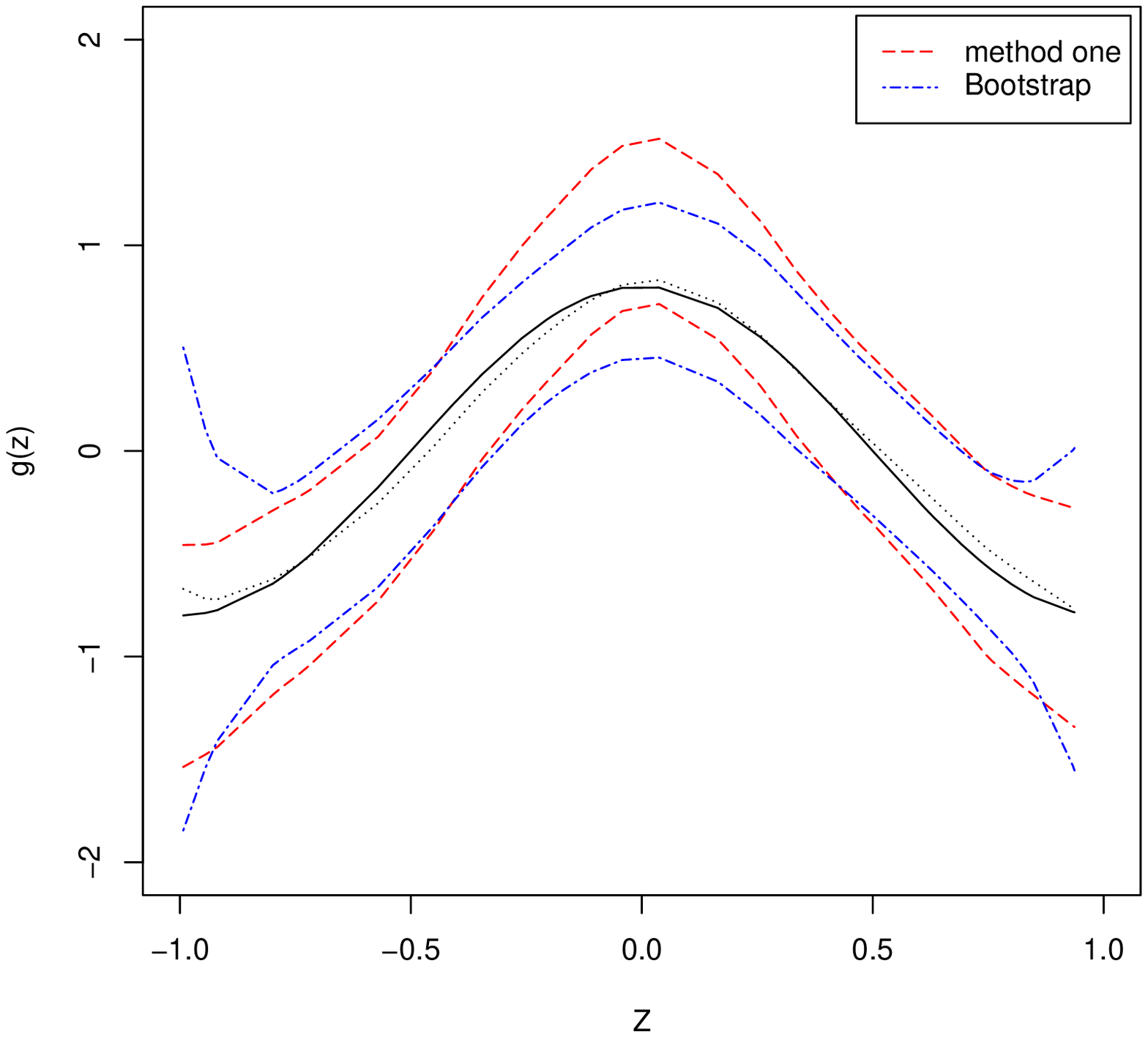}
    \includegraphics[width=4cm,height=6cm]{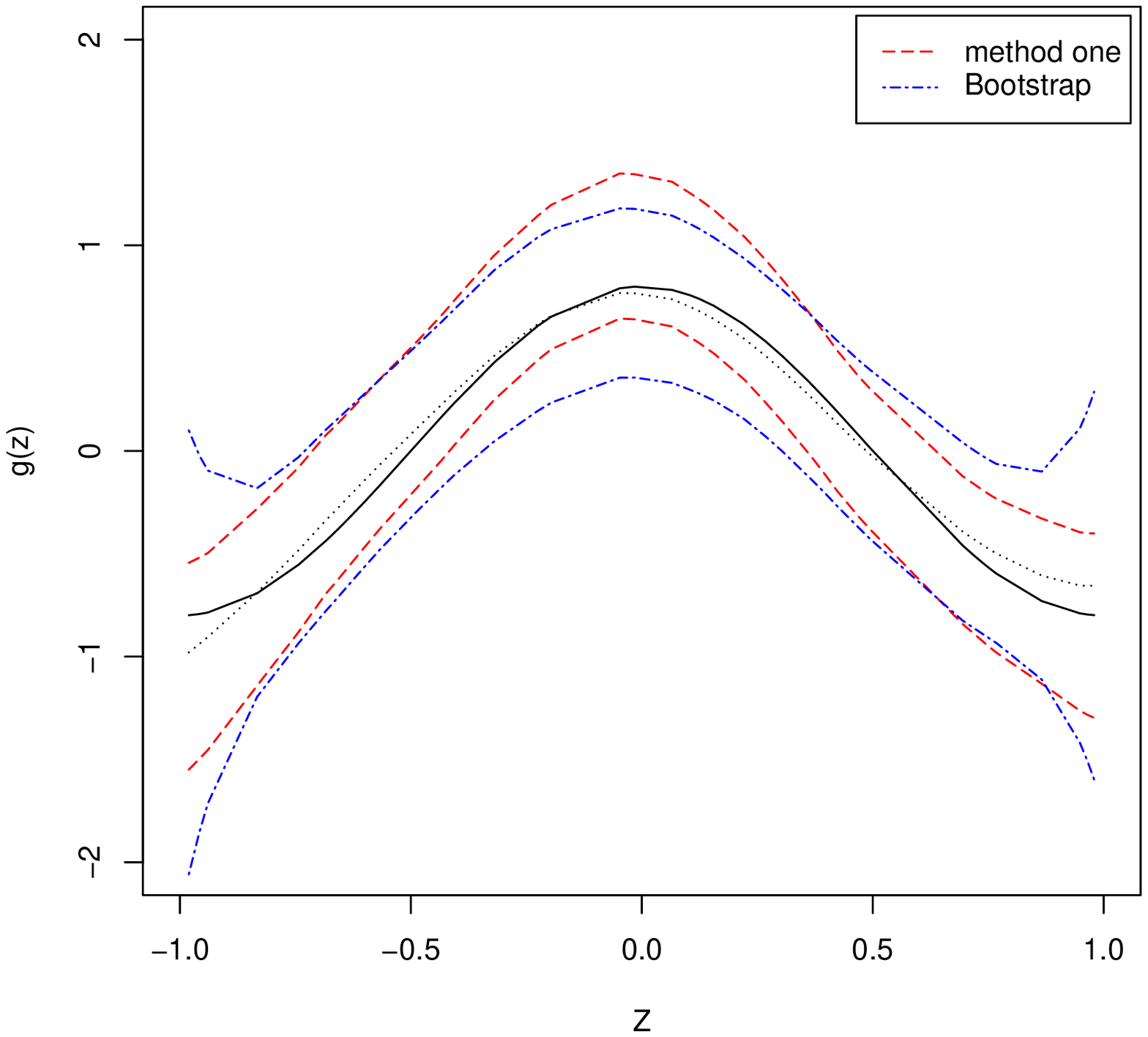}\\
    \includegraphics[width=4cm,height=6cm]{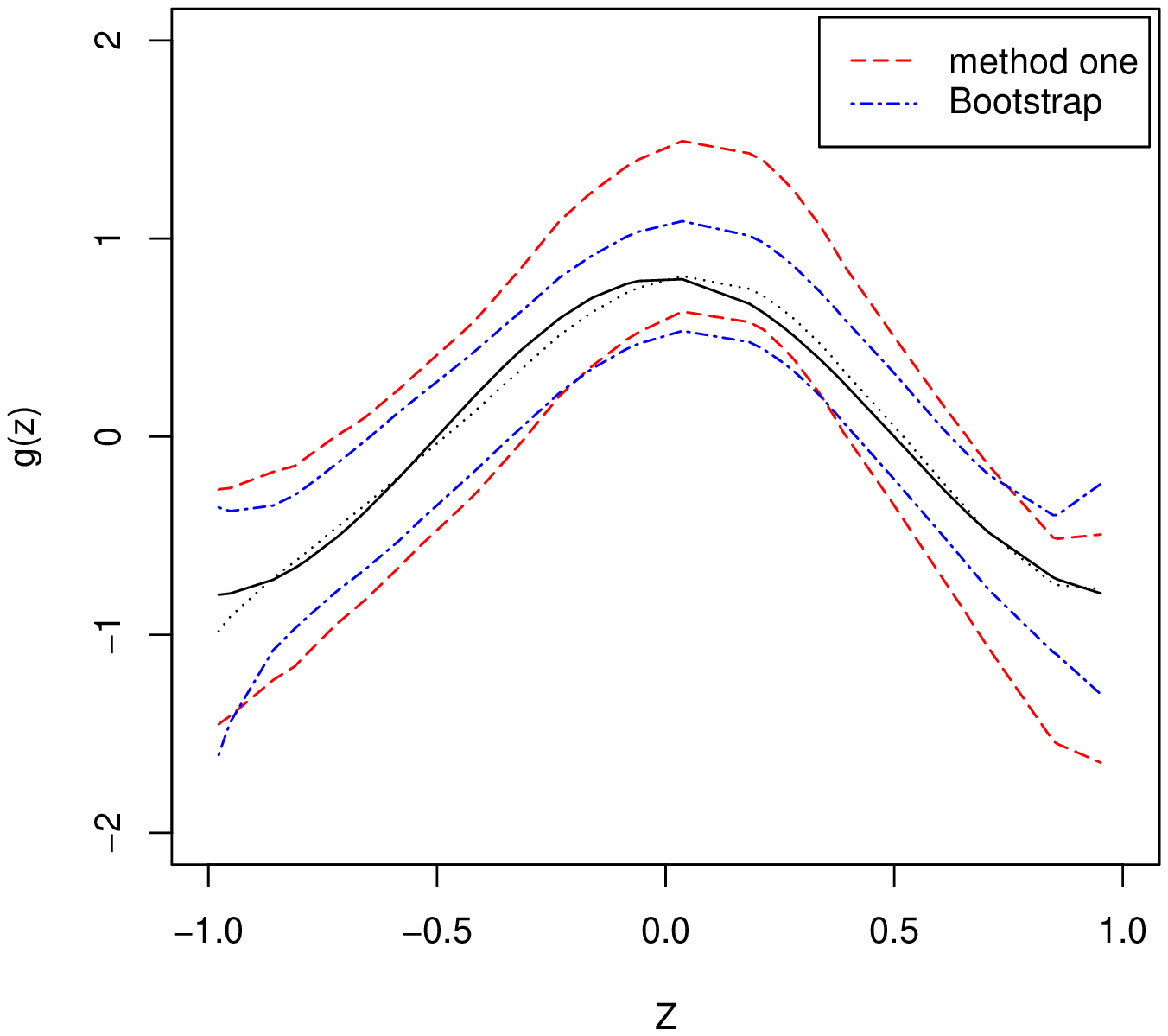}\includegraphics[width=4cm,height=6cm]{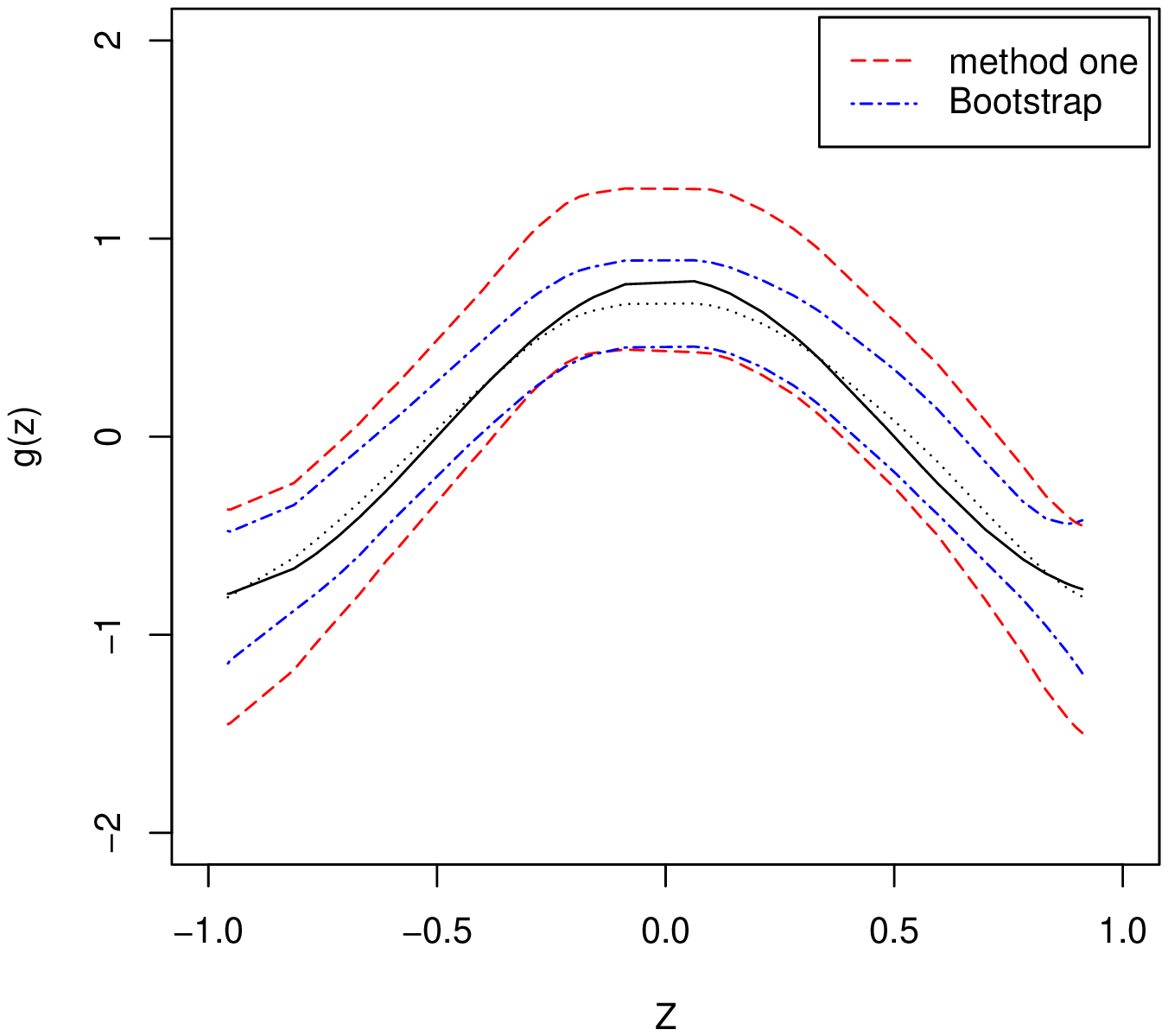}
    \includegraphics[width=4cm,height=6cm]{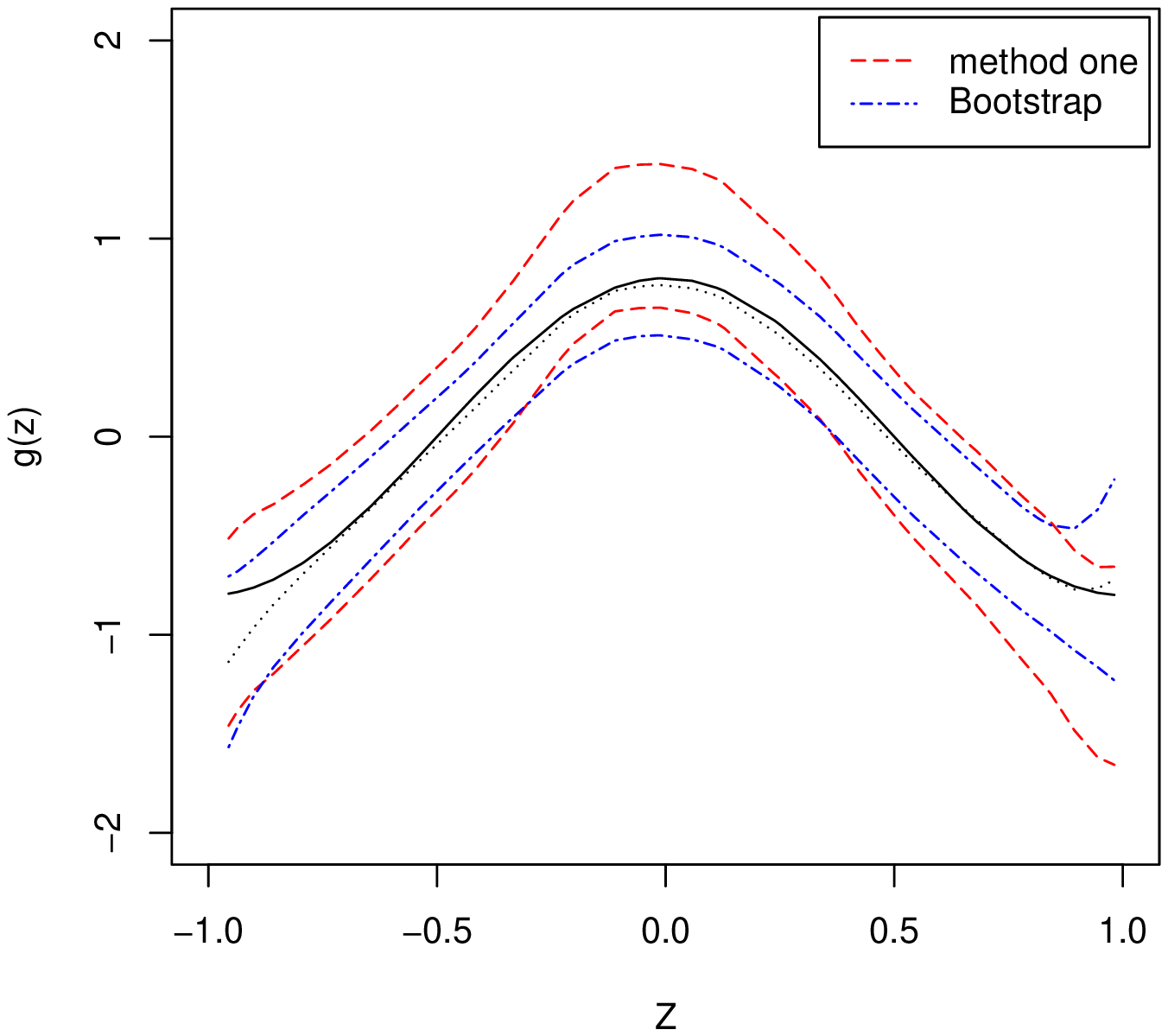}\\
    \includegraphics[width=4cm,height=6cm]{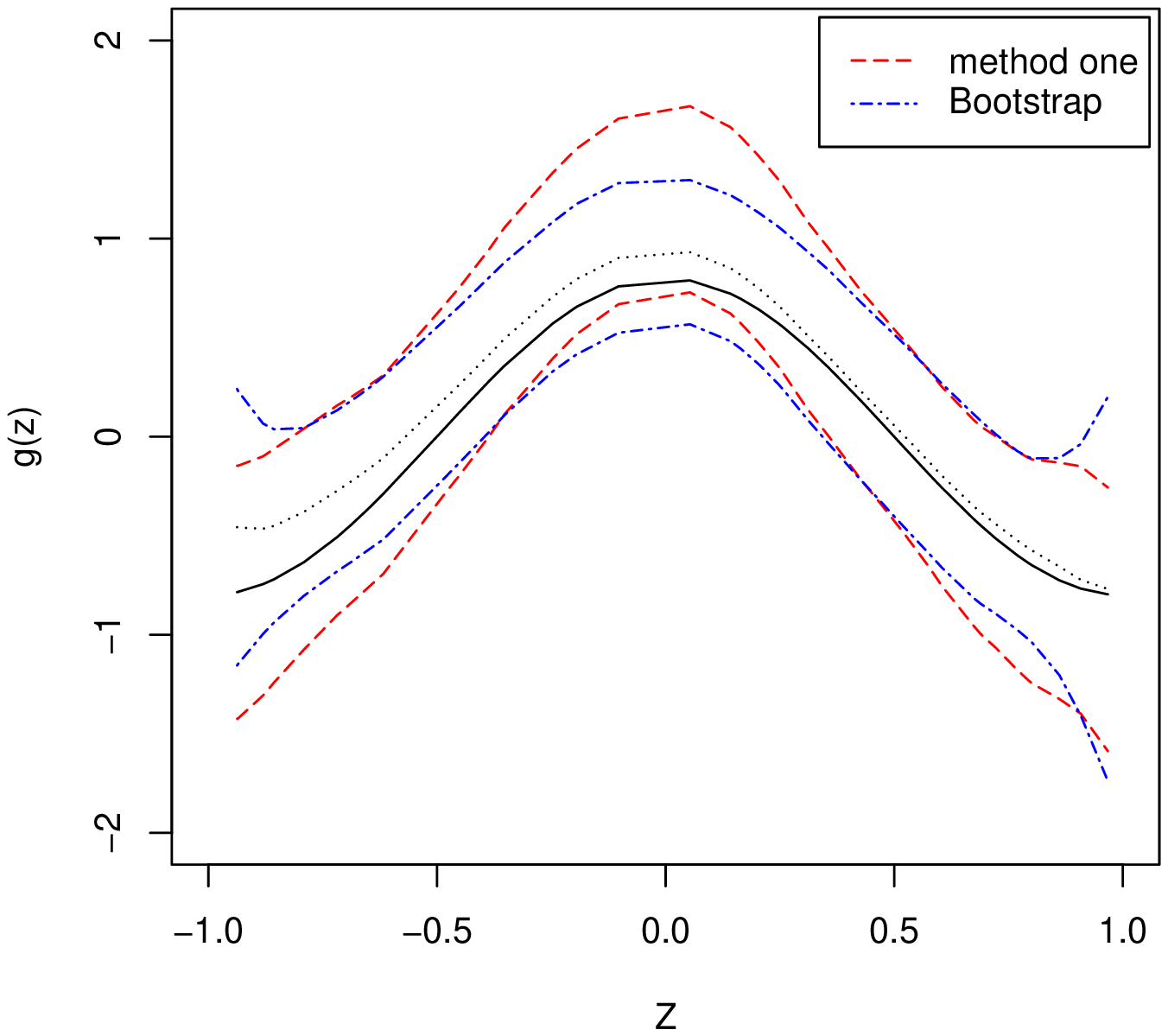}\includegraphics[width=4cm,height=6cm]{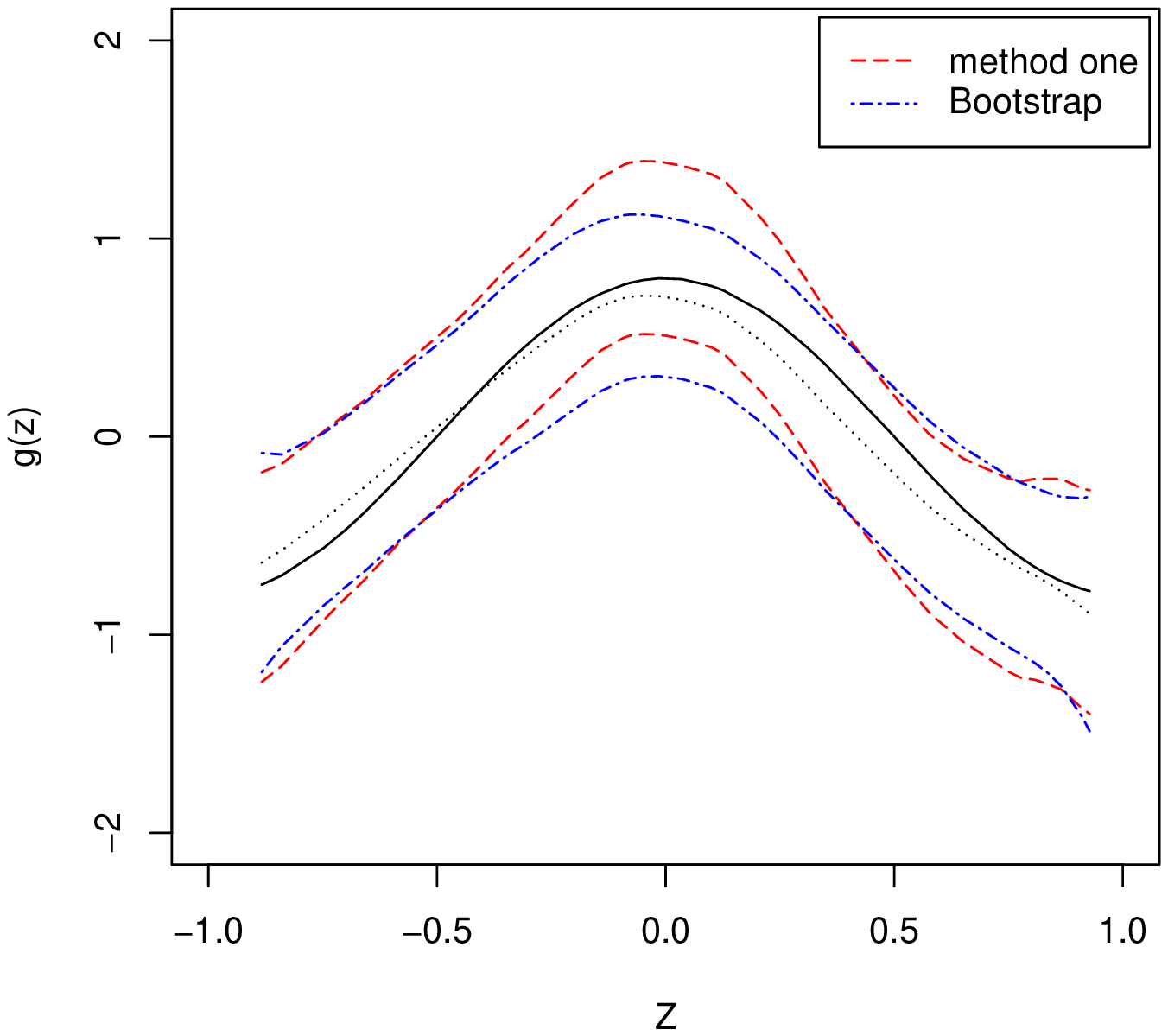}
    \includegraphics[width=4cm,height=6cm]{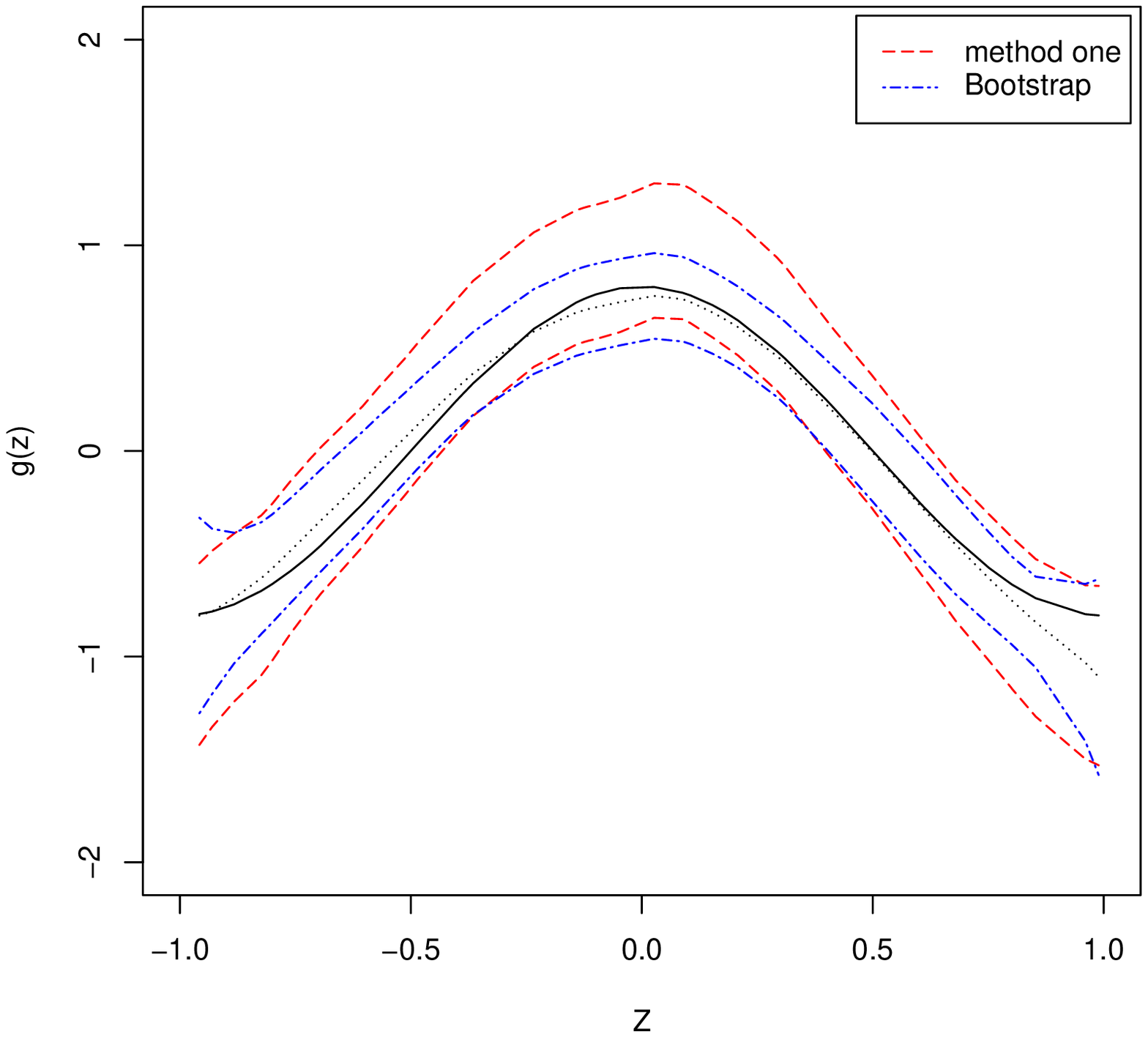}
\caption{{\it The solid lines denote the true curve, the dotted
lines denote the estimated  curve, and the long-dashed lines denote
 the 95\%
 simultaneous confidence bands based on the asymptotic distribution and the dash-dotted lines denote the 95\%
 simultaneous confidence bands based on the Bootstrap procedure for $g(\cdot)$, where figures are displayed for $c=0,0.5,1$ from top to bottom and  for the sample
 sizes
$n=100,150,200$ from left to right, respectively.} }
   \label{fig.1}
\end{figure}

We fix $T=5$ and examine the finite sample performance of the
proposed method when the sample size is taken as $n=100,150$ and
$200$. For each case,  1000 replicates of simulated realizations are
generated, and  the nominal level is taken as $1-\alpha= 0.95$. The
results are given in Tables \ref{tab1}--\ref{tab2} and Figure
\ref{fig.1}. Table \ref{tab1} gives the bias, the standard deviation
and the mean squared error of the estimator $\hat{\bm\beta}$ for
$c=0$ and $c=1$. From Table \ref{tab1}, we can find that the bias,
the standard deviation and the mean squared error are decreased as
the sample size $n$ increases for two cases. For the same sample
size $n$, the results of $c=1$ are better than those of $c=0$. Model
(\ref{3-14}) is reduced to partially linear random effects model
when $c=0$. From (\ref{3-10}) and (\ref{3-13}), it is easy to see
that, in order to remove the fixed effects from the model, we loss
  some sample information to obtain the estimators of parametric and nonparametric
  components.
So the profile least-squares dummy-variable method is not suitable
for the partially linear random effects model, and the resulting
estimators of parametric and nonparametric components are not
efficient. Thus, we need develop the effective estimation procedure
to estimate the random effects models, such as the generalized
profile least squares method or the generalized estimating equation
(GEE).

\begin{table}[!htb]
 \begin{center}
\caption{\label{tab1}  The bias, standard deviation (SD) and mean
squared error (MSE) of $\hat{\bm\beta}$}{ \small {
\begin{tabular}{ ccccccccccc }
\ML & \multicolumn{3}{c}{$c=0$}
 &\multicolumn{3}{c}{$c=1$}\NN
  \cmidrule(r){3-5}\cmidrule(r){6-8}
$\hat{\bm\beta}$ &     & 100         & 150        & 200       & 100         & 150        & 200\\
\hline
              & Bias   &0.0063   &0.0059   &0.0048   &0.0045  &0.0046  &0.0023  \\
$\hat{\beta}_{1}$& SD  &0.0859   &0.0720   &0.0682   &0.0841  &0.0647  &0.0635  \\
              & MSE    &0.0074   &0.0052   &0.0046   &0.0071  &0.0042  &0.0040  \\
\hline
              & Bias   &0.0057   &0.0046   &0.0031   &0.0053  &0.0027  &0.0022  \\
$\hat{\beta}_{2}$& SD  &0.0901   &0.0696   &0.0620   &0.0906  &0.0687  &0.0601  \\
              & MSE    &0.0081   &0.0049   &0.0038   &0.0082  &0.0048  &0.0036  \\
\hline
              & Bias   &0.0062   &0.0049   &0.0042   &0.0041  &0.0029  &0.0026  \\
$\hat{\beta}_{3}$ & SD &0.0912   &0.0770   &0.0650   &0.0857  &0.0679  &0.0545  \\
              & MSE    &0.0083   &0.0059   &0.0042   &0.0074  &0.0046  &0.0031  \\
\hline
\end{tabular}%
}  }
\end{center}
\end{table}
\par

Based on the asymptotic distribution and the Bootstrap method, Table
\ref{tab2} gives the average probabilities of the simultaneous
confidence band  for the nonparametric function $g(\cdot)$ when the
nominal level is $1-\alpha= 0.95$, where ``method one" denotes the
method based on asymptotic distribution and ``Bootstrap" denotes the
method based on the Bootstrap procedure in Table \ref{tab2}. For the
bootstrap procedure, we use $M = 200$ bootstrap replications to
estimate $c_{\alpha}$ and $\mathrm{Var}(\hat{g}(z|\mathcal{D}))$.

\vspace{0.2cm}
\begin{table}[!htb]
 \begin{center}
\caption{\label{tab2}\small  Coverage probabilities of nonparametric
component with the nominal level 95\%} {\small {
\renewcommand\arraystretch{1}
\tabcolsep 9pt \vskip0.5cm
\begin{tabular}{ ccccc }
\hline
 & $n$  &  $c=0$  & $c=0.5$     &$c=1$         \\
\hline
          & 100   &  0.926     &  0.933   &  0.941   \\
method one& 150   &  0.933     &  0.940   &  0.949   \\
          & 200   &  0.946     &  0.951   &  0.953   \\
\hline
          & 100   &   0.928    &  0.934   &    0.942  \\
Bootstrap & 150   &   0.937    &  0.946   &    0.950  \\
          & 200   &   0.948    &  0.952   &    0.954  \\
\hline
\end{tabular}%
} }
\end{center}
\end{table}

\par

From Table \ref{tab2}, it is easy to see that the average coverage
probabilities of the simultaneous confidence band for the
nonparametric function obtained by the two methods tend to 0.95 as
the sample size $n$ increases for three cases. When $c = 0$, the
average coverage probabilities are lower than those of $c = 0.5$ and
$1$. In addition, we also can find that the average coverage
probabilities based on the asymptotic distribution is lower than
those of the Bootstrap method, which implies that the Bootstrap
method performs better than the asymptotic distribution method. The
reason is that the Bootstrap method avoids estimating the asymptotic
bias and variance and reduces the computational burden and
accumulative errors.

Based on the asymptotic distribution and the Bootstrap method,
Figure \ref{fig.1} gives the  95\% pointwise confidence bands of
$g(\cdot)$ for $n = 100, 150, 200$ and $c = 0, 0.5, 1$,
respectively. Figure \ref{fig.1}
 reveals that the performance of asymptotic confidence
bands is not worse than that based on the bootstrap procedure. In
addition, the confidence bands obtained by the two methods become
narrow as the sample size $n$ increases for three cases. From Table
\ref{tab2} and Figure \ref{fig.1}, it is easy to observe that,
although the bootstrap method works better than the method based on
asymptotic distribution, the proposed asymptotic distribution method
is comparable with the bootstrap method.

\section{\bf Appendix:   proofs of the main results}

\par
Let
$\mathbf{P}=(\mathbf{I}_{nT}-\mathbf{M})^{\tau}(\mathbf{I}_{nT}-\mathbf{M})$
and
$\Phi=\sum_{t=1}^{T}\sum_{s=1}^{T}E\{(\widetilde{\bm{X}}_{it}[\widetilde{\bm{X}}_{is}-\sum_{l}\widetilde{\bm{X}}_{il}/T]^{\tau}V_{it}V_{is}\}$.
The following Lemmas  \ref{lema1}--\ref{lema5} play a very important
role in proving the  main results of Theorems
\ref{theo1}--\ref{theo4}, and the details of proofs can be found in
Su and Ullah (2006) and Zhang et al. (2011), we omit the details
here.

{\lemma\label{lema1} Assume that conditions (C1)--(C6) hold. Let $C$
be a positive constant and
$m(Z_{it},z)=\bm{e}^{\tau}(\mathbf{Z}_{z}^{\tau}\mathbf{W}_{z}\mathbf{Z}_{z})^{-1}Z_{z_{it}}K_{h}(Z_{it}-z)$,
where $Z_{z_{it}}$ is a typical column of ${\bf Z}_{z}$, we have

{\rm (i)}~~
$m(Z_{it},z)=n^{-1}K_{h}(Z_{it}-z)f^{-1}(z)\{1+o_{p}(1)\}$,
 where $f(z)=\sum_{t=1}^{T}f_{t}(z)$;

{\rm (ii)}~~
$\lim\limits_{n\rightarrow\infty}P_{n}\Big\{\sup\limits_{z\in[0,1]}\max\limits_{1\leq
i\leq n,1\leq t\leq T}|m(Z_{it},z)|\leq C(nh)^{-1}\Big\}=1.$

}

{\lemma\label{lema2} Assume that conditions (C1)--(C6) hold, we have
$$(\mathbf{D^{\tau}PD})^{-1}=(\mathbf{D^{\tau}D})^{-1}+O_{p}(\zeta_{n})=T^{-1}\mathbf{I}_{n-1}+O_{p}(\zeta_{n}),$$
where $\zeta_{n}=(\bm{e}_{n-1}\bm{e}^{\tau}_{n-1})(nh)^{-1}\sqrt{\ln
n}$. }

{\lemma\label{lema3} Assume that conditions (C1)--(C6) hold, we have

${\rm (i)}
 ~~\frac{1}{n}\mathbf{X^{\tau}PX}\stackrel{P}\longrightarrow\sum_{t=1}^{T}E[(\bm{X}_{it}-\bm{p}(Z_{it}))(\bm{X}_{it}-\bm{p}(Z_{it}))^{\tau}],$

${\rm(ii)}
~~\frac{1}{n}\mathbf{X^{\tau}PD(D^{\tau}D)^{-1}D^{\tau}PX}
\stackrel{P}\longrightarrow
\frac{1}{T}\sum_{t=1}^{T}\sum_{s=1}^{T}E[(\bm{X}_{it}-\bm{p}(Z_{it}))(\bm{X}_{is}-\bm{p}(Z_{is}))^{\tau}],$

${\rm (iii)} ~~
\frac{1}{n}\mathbf{\widetilde{X}^{\tau}\widetilde{Q}\widetilde{X}}\stackrel{P}\longrightarrow\Phi.$
}

{\lemma\label{lema4} Assume that conditions (C1)--(C6) hold, we have
$$\frac{1}{\sqrt{n}}\mathbf{\widetilde{X}^{\tau}\widetilde{Q}}(\mathbf{I}_{n}-\mathbf{M})\bm{g(Z)}=o_{p}(1).$$
}

{\lemma\label{lema5} Assume that conditions (C1)--(C6) hold, we have
\begin{eqnarray*}
&&\frac{1}{\sqrt{n}}\mathbf{X^{\tau}P}\bm{V}=\frac{1}{\sqrt{n}}\sum_{i=1}^{n}\sum_{t=1}^{T}(\bm{X}_{it}-\bm{p}(Z_{it}))V_{it}+o_{p}(1),\\
&&\frac{1}{\sqrt{n}}\mathbf{X^{\tau}PD(D^{\tau}D)^{-1}D^{\tau}P}\bm{V}=\frac{1}{\sqrt{n}T}\sum_{i=1}^{n}\sum_{t=1}^{T}\sum_{s=1}^{T}(\bm{X}_{it}-\bm{p}(Z_{it}))V_{is}+o_{p}(1).
 \end{eqnarray*}
}

\noindent{\bf Proof of Theorem \ref{theo1}.}  The proofs of Theorem
\ref{theo1} can immediately be obtained from Su and Ullah (2006) and
Zhang et al. (2011) by Lemmas \ref{lema1}--\ref{lema5}. So we omit
the details here. \hfill$\Box$

\noindent{\bf Proof of Theorem \ref{theo2}.} Note that
$(\mathbf{I}_{nT}-\mathbf{D(D^{\tau}PD)}^{-1}\mathbf{D^{\tau}P})\mathbf{D}\bm{\alpha}=0$.
By (\ref{3-11}), (\ref{3-13}) and Lemma \ref{lema2}, we have
\begin{eqnarray}\nonumber
\hat{g}(z)&=&\bm{m}^{\tau}(z)(\bm{Y}-\mathbf{D}\bm{\hat{\alpha}}-\mathbf{X}\bm{\hat{\beta}})\\\nonumber
               &=&\bm{m}^{\tau}(z)(\mathbf{I}_{nT}-\mathbf{D(D^{\tau}PD)}^{-1}\mathbf{D^{\tau}P})(\bm{Y}-\mathbf{X}\bm{\hat{\beta}}) \\\nonumber
               &=&\bm{m}^{\tau}(z)(\mathbf{I}_{nT}-\mathbf{D(D^{\tau}PD)}^{-1}\mathbf{D^{\tau}P})(\bm{g}+\bm{V}-\mathbf{X}(\bm{\hat{\beta}-\beta}))\\
               &=&\bm{m}^{\tau}(z)\mathbf{Q}_{1}(\bm{g}+\bm{V}-\mathbf{X}(\bm{\hat{\beta}-\beta})).\label{3-16}
\end{eqnarray}
Invoking the  Taylor expansion, we have
\begin{eqnarray}\label{3-17}
g(Z_{it})\approx
g(z)+g'(z)(Z_{it}-z)+\frac{1}{2}g''(z)(Z_{it}-z)^{2},
\end{eqnarray}
where $Z_{it}$ is close to $z\in[0,1]$. By (\ref{3-16}) and
(\ref{3-17}), we have
\begin{eqnarray}\nonumber
\hat{g}(z)&\approx&\bm{m}^{\tau}(z)(\mathbf{I}_{nT}-\mathbf{D(D^{\tau}PD)}^{-1}\mathbf{D^{\tau}P})g(z)\bm{e}_{nT}+\bm{m}^{\tau}(z)\mathbf{Q}_{1}g'(z)\bm{Z}_{z}\\\nonumber
&&+\frac{1}{2}\bm{m}^{\tau}(z)\mathbf{Q}_{1}g''(z)\bm{Z}_{z}^{2}+\bm{m}^{\tau}(z)\mathbf{Q}_{1}\bm{V}-\bm{m}^{\tau}(z)\mathbf{Q}_{1}\mathbf{X}(\bm{\hat{\beta}-\beta})\\\nonumber
&=&\bm{m}^{\tau}(z)\mathbf{I}_{nT}g(z)\bm{e}_{nT}-\bm{m}^{\tau}(z)\mathbf{D(D^{\tau}PD)}^{-1}\mathbf{D^{\tau}P}g(z)\bm{e}_{nT}+\bm{m}^{\tau}(z)\mathbf{Q}_{1}g'(z)\bm{Z}_{z}\\
&&+\frac{1}{2}\bm{m}^{\tau}(z)\mathbf{Q}_{1}g''(z)\bm{Z}_{z}^{2}+\bm{m}^{\tau}(z)\mathbf{Q}_{1}\bm{V}-\bm{m}^{\tau}(z)\mathbf{Q}_{1}\mathbf{X}(\bm{\hat{\beta}-\beta})
,\label{taile}
\end{eqnarray}
where
$\bm{Z}_{z}=(Z_{11}-z,\cdots,Z_{1T}-z,Z_{21}-z,\cdots,Z_{2T}-z,\cdots,Z_{n1}-z,\cdots,Z_{nT}-z)^{\tau}$.
For ease of notation, let
$S_{nT,l}(z)=\sum_{i=1}^{n}\sum_{t=1}^{T}K_{h}(Z_{it}-z)(Z_{it}-z)^{l},~~l=0,
1, 2.$ For the first term  of (\ref{taile}), some simple
calculations yield that
\begin{eqnarray}\nonumber
\bm{m}^{\tau}(z)\mathbf{I}_{nT}g(z)\bm{e}_{nT}
&=&(1,0)(\mathbf{Z}_{z}^{\tau}\mathbf{W}_{z}\mathbf{Z}_{z})^{-1}\mathbf{Z}_{z}^{\tau}\mathbf{W}_{z}\mathbf{I}_{nT}\bm{e}_{nT}g(z)\\\nonumber
&=&(1,0)\left (\begin{array}{cc} S_{nT,0}(z)& S_{nT,1}(z)\\
S_{nT,1}(z) & S_{nT,2}(z) \\
\end{array} \right )^{-1}
\left
(\begin{array}{c}S_{nT,0}(z)\\
S_{nT,1}(z)\\
\end{array} \right )g(z)\\\nonumber
&=&(1,0)\left (\begin{array}{cc}S_{nT,2}(z)& -S_{nT,1}(z)\\
-S_{nT,1}(z) & S_{nT,0}(z) \\
\end{array} \right )\left
(\begin{array}{c}S_{nT,0}(z)\\
S_{nT,1}(z)\\
\end{array} \right )g(z)\\\nonumber
&&\times\Big(S_{nT,0}(z)S_{nT,2}(z)-S_{nT,1}^{2}(z)\Big)^{-1}\\\nonumber
&=&(1,0)\left
(\begin{array}{c}S_{nT,0}(z)S_{nT,2}(z)-S_{nT,1}^{2}(z)\\
0\\
\end{array} \right )g(z)\\\nonumber
&&\times\Big(S_{nT,0}(z)S_{nT,2}(z)-S_{nT,1}^{2}(z)\Big)^{-1}\\
&=&g(z)\label{taile1}.
\end{eqnarray}
By (\ref{taile}), (\ref{taile1}) and some calculations, we have
\begin{eqnarray}\nonumber
\sqrt{nh}(\hat{g}(z)-g(z))&\approx&
\sqrt{nh}\bm{m}^{\tau}(z)\mathbf{Q}_{1}g'(z)\bm{Z}_{z}+\frac{\sqrt{nh}}{2}\bm{m}^{\tau}(z)\mathbf{Q}_{1}g''(z)\bm{Z}_{z}^{2}+\sqrt{nh}\bm{m}^{\tau}(z)\mathbf{Q}_{1}\bm{V}
\\\nonumber
&&-\sqrt{nh}\mathbf{D(D^{\tau}PD)}^{-1}\mathbf{D^{\tau}P}g(z)\bm{e}_{nT}-\sqrt{nh}\bm{m}^{\tau}(z)\mathbf{Q}_{1}\mathbf{X}(\bm{\hat{\beta}-\beta})\\
&=:&J_{11}+J_{12}+J_{13}-J_{14}-J_{15}.\label{fenjie}
\end{eqnarray}
From the results of Lemmas \ref{lema1}--\ref{lema4}, it is easy to
show that $J_{11}=o_{p}(1)$ and $J_{14}=o_{p}(1)$. Again invoking
the results of Lemmas \ref{lema1}--\ref{lema3} and
$\|\hat{\bm\beta}-\bm\beta\|=O_{p}(n^{-1/2})$ in Theorem
\ref{theo1}, we can prove that $J_{15}=o_{p}(1)$.

Now we consider $J_{12}$ and $J_{13}$. Let $\bm{M}(Z_{it},z)$ be a
typical column of $\mathbf{M}(z)$, where
$\mathbf{M}(z)=(\bm{M}(Z_{11},z),\cdots, \bm{M}(Z_{1T},z),
\bm{M}(Z_{21},z),\cdots,\bm{M}(Z_{2T},z),\cdots,\bm{M}(Z_{n1},z),\cdots,\bm{M}(Z_{nT},z))$.
For $J_{12}$, by Lemma \ref{lema1} and some calculations, we can
show that
\begin{eqnarray}\nonumber
J_{12}&\approx&\frac{\sqrt{nh}}{2}\sum_{i=1}^{n}\sum_{t=1}^{T}(1,0)\bm{M}(Z_{it},z)g''(z)(Z_{it}-z)^{2}\\\nonumber
&=&\frac{\sqrt{nh}}{2}\frac{1}{nf(z)}\sum_{i=1}^{n}\sum_{t=1}^{T}K_{h}(Z_{it}-z)g''(z)(Z_{it}-z)^{2}+o_{p}(1)\\\nonumber
&=&\frac{\sqrt{nh}}{2}\frac{1}{nf(z)}g''(z)\int
z^{2}K(z)dz+o_{p}(h^{2})\\
&=&\frac{\sqrt{nh}}{2}b(z)+o_{p}(h^{2}).\label{j12}
\end{eqnarray}
By Lemma \ref{lema2} and Lemma \ref{lema5}, and using the same
argument for $J_{13}$ and some simple calculations, we can show that
\begin{eqnarray}\nonumber
J_{13}&=&\sqrt{nh}\bm{m}^{\tau}(z)\mathbf{Q}_{1}{\bm V}\\\nonumber
&=&\sqrt{nh}\frac{1}{nf(z)}\sum_{i=1}^{n}\sum_{t=1}^{T}K_{h}(Z_{it}-z)\widetilde{V}_{it}+o_{p}(1)\\
&\stackrel{L}\longrightarrow& N(0,\Sigma_{g}),\label{j13}
\end{eqnarray}
where
$\widetilde{V}_{it}={V}_{it}-\frac{1}{T}\sum\limits_{s=1}^{T}{V}_{is}$
and $\Sigma_{g}=\nu_{0}\bar{\sigma}^{2}(z)f^{-2}(z)$.
\par
By (\ref{fenjie}) and (\ref{j12}), it is easy to obtain that
\begin{eqnarray}\nonumber
\hat{g}(z)-g(z)-b(z)&=&{{\bm
m}^{\tau}}(z)(\mathbf{I}_{nT}-\mathbf{D(D^{\tau}PD)}^{-1}\mathbf{D^{\tau}P})\bm{V}+o_{p}(1)\\\nonumber
&\approx&\bm{m^{\tau}}(z)\bm{\widetilde{V}}+o_{p}(1)\\\nonumber
&=&(1,0)(\mathbf{Z}_{z}^{\tau}\mathbf{W}_{z}\mathbf{Z}_{z})^{-1}\mathbf{Z}_{z}^{\tau}\mathbf{W}_{z}\bm{\widetilde{V}}+o_{p}(1)\\
&=:&I_{1}(z)+o_{p}(1),\label{3-18}
\end{eqnarray}
where
$\widetilde{\bm{V}}=(\widetilde{V}_{11},\cdots,\widetilde{V}_{1T},\widetilde{V}_{21},\cdots,\widetilde{V}_{2T},\cdots,\widetilde{V}_{n1},\cdots,\widetilde{V}_{nT})^{\tau}$
and $\widetilde{V}_{it}=V_{it}-\frac{1}{T}\sum_{s=1}^{T}V_{is}$.
\par
Next, we approximate the process $I_{1}(z)$ as follows. Note that
\begin{eqnarray*}\nonumber
 \mathbf{Z}_{z}^{\tau}\mathbf{W}_{z}\mathbf{Z}_{z}
=\left
(\begin{array}{cc}\displaystyle\sum_{i=1}^{n}\sum_{t=1}^{T}\displaystyle\
K_{h}(Z_{it}-z) &
\displaystyle\sum_{i=1}^{n}\sum_{t=1}^{T}\displaystyle
K_{h}(Z_{it}-z)(Z_{it}-z)\\
\displaystyle\sum_{i=1}^{n}\sum_{t=1}^{T}\displaystyle
K_{h}(Z_{it}-z)(Z_{it}-z) &
\displaystyle\sum_{i=1}^{n}\sum_{t=1}^{T}\displaystyle K_{h}(Z_{it}-z)(Z_{it}-z)^{2} \\
\end{array} \right ).
\end{eqnarray*}
By Lemma \ref{lema1}, we have
\begin{eqnarray}\label{3-19}
n\mathbf{H}(\mathbf{Z}_{z}^{\tau}\mathbf{W}_{z}\mathbf{Z}_{z})^{-1}\mathbf{H}=f^{-1}(z)\Omega^{-1}+O_{p}(h+(\log
n/nh)^{1/2}),
\end{eqnarray}
where $\mathbf{H}=\left (\begin{array}{cc}1 &0\\
0 & h
\end{array} \right )$ and
$\Omega=\left (\begin{array}{cc}1 &0\\
0 & \mu_{2}
\end{array} \right )$.\\
By Lemma \ref{lema1}, we further obtain that
\begin{eqnarray}\label{3-20}
\left\|\frac{1}{n}\mathbf{H}^{-1}\mathbf{Z}_{z}^{\tau}\mathbf{W}_{z}\bm{\widetilde{V}}\right\|_{\infty}=O_{p}(h+(\log
n/nh)^{1/2}).
\end{eqnarray}
By (\ref{3-19}) and (\ref{3-20}), we have
\begin{eqnarray}\label{3-21}
\Big\|I_{1}(z)-\frac{1}{nf(z)}(1,0)\Omega^{-1}\mathbf{H}^{-1}\mathbf{Z}_{z}^{\tau}\mathbf{W}_{z}\bm{\widetilde{V}}\Big\|_{\infty}
=O_{p}\Big(h(\log n/nh)^{1/2}+(\log n/nh)\Big).
\end{eqnarray}
Let
\begin{eqnarray*}\nonumber
I_{2}(z)&=:&\frac{1}{nf(z)}(1,0)\Omega^{-1}\mathbf{H}^{-1}\mathbf{Z}_{z}^{\tau}\mathbf{W}_{z}\bm{\widetilde{V}}\\
&=&\frac{1}{nf(z)}\sum_{i=1}^{n}\sum_{t=1}^{T}K_{h}(Z_{it}-z)\widetilde{V}_{it}.
\end{eqnarray*}
Invoking Theorem 1 and Lemma 1 in Fan and Zhang (2000), for
$h=n^{-\rho}$, $1/5\leq \rho\leq 1/3$, we have
\begin{eqnarray}\label{3-22}
P\Big\{(-2\log
h)^{1/2}\left(\left\|(nh\Sigma_{g}^{-1})^{1/2}I_{2}(z)\right\|_{\infty}-d_{n}\right)<u\Big\}
\longrightarrow \exp \left(-2\exp(-u)\right),
\end{eqnarray}
where $\Sigma_{g}=\nu_{0}\bar{\sigma}^{2}(z)f^{-2}(z)$ is defined in
Theorem \ref{theo1} and $d_{n}$ is defined in Theorem \ref{theo2}.
By (\ref{3-20}), (\ref{3-21}) and (\ref{3-22}), we complete the
proof of Theorem \ref{theo2}. \hfill$\Box$
\par

\noindent{\bf Proof of Theorem \ref{theo3}.}   Along the same lines
as the proof of Theorem \ref{theo2}, it is easy to prove Theorem
\ref{theo3}. Thus, we omit the details of proof. \hfill$\Box$

\noindent{\bf Proof of Theorem \ref{theo4}.}\ \ To prove Theorem
\ref{theo4}, we need derive the rate of convergence for the bias and
variance estimators. We first consider the difference between
$\mathrm{bias}(\hat{g}(z))$ and
$b(z)=\frac{1}{2}h^{2}\mu_{2}g''(z)$. By (\ref{3-19}) and its
similar arguments, we have
\begin{equation}\label{3-23}
\Big\|\widehat{\mathrm{bias}}(\hat{g}(z)|\mathcal{D})-b(z)\Big\|_{\infty}=O_{p}(h^{2}\{\sqrt{\log
n/nh^{5}_{*}}\})=O_p\Big(h^2(n^{-1/7}\log^{1/2}n)\Big),
\end{equation}
where $h_*=O(n^{-1/7})$.

Furthermore, by Lemmas \ref{lema1}--\ref{lema2}, and similar
argument of (\ref{3-20}), we have
$$\left\|\frac{h}{n}\mathbf{H}^{-1}(\mathbf{Z}_{z}^{\tau}
\mathbf{W}_{z}\mathbf{Q}_{1}\mathbf{W}_{z}\mathbf{Z}_{z})\mathbf{H}^{-1}-f(z)\Lambda\right\|_{\infty}=o_p(1),$$
where $\Lambda=\left (\begin{array}{cc}\nu_{0} &0\\
    0 & \nu_{2}
            \end{array} \right )$.
By the similar argument, it is easy to check that
$\Big\|\hat{\sigma}^{2}(z)-\sigma^{2}(z)\Big\|_{\infty}=o_{p}(1)$.
These results, together with Theorem \ref{theo2}, we can show that,
uniformly for $z\in[0,1]$,
\begin{equation}\label{3-24}
\Big\|nh\widehat{\mathrm{Var}}\{\hat{g}(z)|\mathcal{D}\}-\Sigma_{g}\Big\|_{\infty}=o_p(1).
\end{equation}
By (\ref{3-23}) and (\ref{3-24}), and invoking the result of Theorem
\ref{theo2}, we finish the proof of Theorem \ref{theo4}.\hfill$\Box$

%

\bibliographystyle{unsrt} 

\end{document}